\documentclass[showpacs,preprintnumbers,amssymb]{revtex4-1}

\usepackage[intlimits]{amsmath}

\usepackage{graphicx}
\usepackage{dcolumn}
\usepackage{bm}

\usepackage{comment}

\usepackage[sort&compress]{natbib}
\usepackage[mathscr]{eucal}
\usepackage{color}
\definecolor{DarkBlue}{rgb}{0.1,0.1,0.5}
\definecolor{Red}{rgb}{0.9,0.0,0.1}
\definecolor{Grey}{rgb}{0.7,0.7,0.7}

\newcommand{\beq}{\begin{equation}}
\newcommand{\eeq}{\end{equation}}
\newcommand{\bey}{\begin{eqnarray}}
\newcommand{\eey}{\end{eqnarray}}

\newcommand{\dt}{\partial_t}
\newcommand{\bfk}{\mathbf{k}}

\newcommand{\nmk}{\mathrm{k}}

\newcommand{\UDJ}[3]{\widehat{\tilde{u}}_{#1}(\mathbf{#2},{#3})}
\newcommand{\pauli}{\sigma_{\scriptscriptstyle{2}}}
\newcommand{\paulitwo}[1]{\big[\sigma_{\scriptscriptstyle{2}}\big]_{\scriptscriptstyle{\text{\textsc{#1}}}}}

\newcommand{\uotu}[3]{U_{\scriptstyle{#1}}(\mathbf{#2};{#3})}
\newcommand{\Gramma}[2]{\mathcal{M}_{\mathbf{#1}}}
\newcommand{\bigG}[2]{G_{\scriptstyle{\mathbf{#1}}}}
\newcommand{\timo}[2]{#1\{#2 #1\}_{\scriptscriptstyle{\mathrm{T}}}}

\newcommand{\spinu}[4]{\mathcal{U}_{\scriptscriptstyle{\text{\textsc{#1}}}}({#2},\mathbf{#3};{#4})}
\newcommand{\spinnu}[2]{\mathcal{U}_{\mathbf{#1}}}

\newcommand{\Fotu}[2]{\mathcal{F}_{\mathbf{#1}}}
\newcommand{\Dotu}[2]{\mathcal{D}_{\mathbf{#1}}}
\newcommand{\Totu}[2]{\mathcal{M}^{\scriptscriptstyle{(0)}}_{\mathbf{#1}}}

\newcommand{\neta}[2]{\eta_{\mathbf{#1}}}

\newcommand{\Ggib}[2]{\big[G_{\mathbf{#1}}\big]^{\mms \scriptscriptstyle{1}}}
\newcommand{\Ggibo}[2]{\big[G^{\scriptscriptstyle{(0)}}_{\mathbf{#1}}\big]^{\mms \scriptscriptstyle{1}}}

\newcommand{\bigma}[2]{\Sigma_{\mathbf{#1}}}
\newcommand{\Rgib}[2]{\big[R_{\mathbf{#1}}\big]^{\mms \scriptscriptstyle{1}}}
\newcommand{\Rgibo}[2]{\big[R^{\scriptscriptstyle{(0)}}_{\mathbf{#1}}\big]^{\mms \scriptscriptstyle{1}}}
\newcommand{\bigR}[2]{R_{\mathbf{#1}}}
\newcommand{\bigRo}[2]{R^{\scriptscriptstyle{(0)}}_{\mathbf{#1}}}

\newcommand{\MOM}[2]{M_{#1}^{\scriptscriptstyle{(0)}}(\mathbf{#2})}

\newcommand{\Bdfnt}[1]{\delta_{\mathbf{#1}}} 

\newcommand{\igl}[1]{\!\int \!\! d^{\scriptstyle{3}}\mathrm{#1}\;} 
\newcommand{\Rigls}[3]{\!\int_{#2}^{#3} \!\!\! d#1} 

\newcommand{\PDEop}[1]{\frac{\partial}{\partial #1}} 

\newcommand{\sms}{\!\frac{\,}{\,}\!} 	
\newcommand{\mms}{\frac{\ }{\ }} 	

\newcommand{\Uuuu}[4]{\tilde{u}_{#1}^{\scriptscriptstyle{#2}}\!(\mathbf{#3},{#4})} 	
\newcommand{\Uout}[2]{\Uuuu{#1}{(0)}{#2}{t}} 					
\newcommand{\Uxuu}[3]{\tilde{u}_{#1}(\mathbf{#2},{#3})}  				
\newcommand{\Uouu}[3]{\tilde{u}_{#1}^{\scriptscriptstyle{(0)}}\!(\mathbf{#2},{#3})}  	

\newcommand{\MTO}[2]{M_{#1}(\mathbf{#2})} 					
\newcommand{\MTOO}[2]{M_{#1}^{\scriptscriptstyle{(0)}}(\mathbf{#2})} 		

\newcommand{\FFF}[3]{\tilde{f}_{#1}(\mathbf{#2},#3)} 					

\newcommand{\Pj}[2]{P_{#1}(\mathbf{#2})} 					

\newcommand{\Rrrr}[4]{R_{#1}^{\scriptscriptstyle{#2}}\!(\mathbf{#3};\:#4)} 	
\newcommand{\Rrxr}[3]{R_{#1}(\mathbf{#2};#3)} 					
\newcommand{\Rror}[3]{\Rrrr{#1}{(0)}{#2}{#3}} 					
\newcommand{\Rrot}[2]{\Rrrr{#1}{(0)}{#2}{t,t'}} 				

\newcommand{\Cccc}[4]{C_{#1}^{\scriptscriptstyle{#2}}(\mathbf{#3};#4)} 		
\newcommand{\Ccxc}[3]{C_{#1}(\mathbf{#2};#3)} 					
\newcommand{\Ccxt}[2]{\Cccc{#1}{\:}{#2}{t,t'}} 					
\newcommand{\Ccot}[2]{\Cccc{#1}{(0)}{#2}{t,t'}} 				

\newcommand{\uu}[2]{\tilde{u}_{\mathbf{#1}}^{\!\scriptscriptstyle{(#2)}}}
\newcommand{\uuo}[1]{\uu{#1}{0}}
\newcommand{\unn}[1]{\tilde{u}_{\mathbf{#1}}}
\newcommand{\mm}[1]{M_{\mathbf{#1}}}
\newcommand{\mmo}[1]{M_{\!\mathbf{#1}}^{\scriptscriptstyle{(0)}}} 		

\newcommand{\rro}[1]{R_{\mathbf{#1}}^{\scriptscriptstyle{(0)}}}
\newcommand{\rrx}[1]{R_{\mathbf{#1}}}

\newcommand{\cco}[1]{C_{\mathbf{#1}}^{\scriptscriptstyle{(0)}}}
\newcommand{\ccx}[1]{C_{\mathbf{#1}}}

\newcommand{\ff}[1]{f_{\mathbf{#1}}}

\newcommand{\dkk}{\da_{\mathbf{k}+\mathbf{k}'}}
\newcommand{\dl}[2]{\da_{\mathbf{#1}\!+\mathbf{#2}}}

\newcommand{\crln}[1]{\langle #1 \rangle} 					
\newcommand{\Crln}[1]{\big\langle #1 \big\rangle} 				
\newcommand{\CRLN}[1]{\Bigg\langle #1 \Bigg\rangle} 				

\newcommand{\bfj}{\mathbf{j}} 							
\newcommand{\nmj}{\mathrm{j}} 							
\newcommand{\bfl}{\mathbf{l}} 							
\newcommand{\nml}{\mathrm{l}} 							

\newcommand{\WCfour}[4]{\crln{#1#2}\crln{#3#4}+\crln{#1#3}\crln{#2#4}+\crln{#2#3}\crln{#1#4}}

\newcommand{\al}{\alpha}
\newcommand{\ba}{\beta}
\newcommand{\ga}{\gamma}
\newcommand{\da}{\delta}
\newcommand{\ea}{\epsilon}

\newcommand{\la}{\lambda}

\newcommand{\om}{\omega}

\newcommand{\order}[1]{\mathcal{O}(\lambda^{#1})}

\begin{document}
\title{Eulerian Field-Theoretic Closure Formalisms for Fluid Turbulence}

\author{Arjun Berera}
\email{ab@ph.ed.ac.uk}
\affiliation{School of Physics and Astronomy, University of Edinburgh,
JCMB, The Kings' Buildings, 
Edinburgh EH9 3JZ, UK}

\author{Matthew Salewski}
\email{matthew.salewski@physik.uni-marburg.de}
\affiliation{Fachbereich Physik, Philipps-Universit\"at Marburg, \\ Renthof 6, Marburg 35037, DE}

\author{W. D. McComb}
\email{wdm@ph.ed.ac.uk}
\affiliation{School of Physics and Astronomy, University of Edinburgh, \\
JCMB, The Kings' Buildings, 
Edinburgh EH9 3JZ, UK}

\date{\today}

\begin{abstract}

The formalisms of Wyld ({\em Annals of Physics}, 14:143, 1961) and 
Martin, Siggia, and Rose (MSR) ({\it Physical Review A}, 8(1):423, July 1973) address the 
closure problem of a statistical treatment of homogeneous isotropic turbulence (HIT) based on techniques primarily developed for quantum field theory. 
In the Wyld
formalism, there is a well-known double-counting problem, for
which an {\it ad hoc} solution was suggested
by Lee ({\it Annals of Physics}, 32:292, (1965)).
We show how to implement this correction in a more
natural way from the basic equations of the formalism.
This leads to what we call the
{\em Improved Wyld-Lee Renormalized Perturbation Theory}.  
MSR had noted that their formalism had more vertex functions than Wyld's
formalism and based on this felt Wyld's formalism was
incorrect.  However a careful comparison of both formalisms
here shows that the Wyld formalism follows a different procedure
to that of the MSR formalism and so the treatment of vertex corrections
appears in different ways in the two formalisms.
Taking that into account, along with
clarifications made to both formalisms,
we find that they are equivalent and we demonstrate this
up to fourth order.
\end{abstract}


In press with {\it Physical Review E}

\maketitle
\section{Introduction}
\subsection{Closure theories and formalisms in turbulence}
In turbulence, statistical closures are techniques 
employed to close the moment-hierarchy established in a statistical treatment 
of the Navier-Stokes equation (NSE). Such theories postulate a relationship 
between high- and low-order moments by way of physical arguments. The goal 
is to accurately describe and predict the statistics of a turbulent system 
while maintaining a strong connection to the underlying dynamics of the NSE. 
Closure theories can, in principle, allow efficient computation of turbulent 
statistics without the computationally intensive demands of Direct Numerical 
Simulation (DNS) \cite{Davidson04}. This feature makes closure theories particularly attractive 
when computational power is limited. Although the ability to compute the 
full Navier-Stokes equations using DNS is currently and increasingly more 
tractable, closure-based computations are still able to provide useful 
insights into turbulent systems at a much smaller computational cost.

The analytic study of turbulence can be classified into different formalisms
from which specific closure theories can be constructed.
Two key formalisms which have developed, and are the focus of this
paper, are the 
Wyld \cite{Wyld61} and
Martin, Siggia, and Rose (MSR) \cite{MaSiRo73} formalisms.
The closure theories themselves that can be
obtained from these formalisms can be further classified 
into two main groups,
Eulerian and Lagrangian, distinguished by the reference frame from which
the dynamics is described.
The former describes a fluid from a `lab', 
where the fluid moves relative to a fixed frame of reference 
outside the fluid. In particular this is the frame
of the Navier-Stokes equations. The Lagrangian 
description is in fact a re-formulation of fluid dynamics that calculates 
statistics by following fluid particles.
The work presented here does not consider 
the Lagrangian closures but instead focuses on those of the standard 
Eulerian formulation.

One of the earliest closures for homogeneous isotropic turbulence
was the quasi-normal approximation.
In this approximation
the fourth-order correlation is written in terms of products of 
second-order correlations \cite{Millionshchikov41,PrRe54}. However
turbulence has highly non-Gaussian correlations 
and so the resultant predictions from this simple approximation
were inadequate, in fact leading to 
total kinetic energy having negative values \cite{Ogura63}. The failure 
of quasi-normality triggered research that made improvements upon it, 
resulting in the EDQNM-methods (see for example \cite{Lesieur90}),
which are a subset of the more general class of
Renormalized 
Perturbation Theories (RPTs). The RPT approach
re-sums certain selected infinite terms from the conventional
perturbation expansion. The underlying idea is these resummations
can capture essential non-perturbative physics of turbulence.
There is a body of literature that has grown up around this work and some 
principle sources elaborating these directions are the books by
Leslie \cite{Leslie73} and McComb \cite{McComb90}.

\subsection{Equations and quantities}
\label{equations}
The main focus of the theories described above is to compute the two-point, two-time correlation function, $\Crln{u_{\al}(\mathbf{x},t)u_{\ba}(\mathbf{x'},t')}$, of a homogeneous and isotropic field of incompressible fluid turbulence. Although this quantity is defined in real-space, we will find it easier to start with the spectral Navier-Stokes equation (NSE),
\beq\label{Spec_NSE}
\Big(\dt+\nu\nmk^2\Big) \Uxuu{\al}{k}{t}  =  \FFF{\al}{k}{t} + \MOM{\al\ba\ga}{k}\!\!\!\!\int_{\bfj+\bfl=\bfk}\!\!\!\!\!d^3\nmj \,d^3\nml\,\Uxuu{\ba}{j}{t}\Uxuu{\ga}{l}{t}
\eeq
defined using the following Fourier convention 
\begin{eqnarray}
a_{\alpha}(\mathbf{x},t) = \int \!\! d\bfk \, \tilde{a}_{\alpha}(\mathbf{k},t)e^{i\mathbf{k\cdot x}} \,, &\qquad& \tilde{a}_{\alpha}(\mathbf{k},t) = \frac{1}{(2\pi)^3}\int \!\! d\mathbf{x} \, a_{\alpha}(\mathbf{x},t)e^{-i\mathbf{k\cdot x}} \,, \\
&\qquad& \nonumber
\end{eqnarray}
where, for the velocity field $a=u$ and for the forcing $a=f$. 
The notation $\bfj+\bfl=\bfk$ under the integral signs indicates that the integration variables are constrained under convolution. Note that the pressure term has been removed using the continuity equation, $\bfk\cdot\mathbf{\tilde{u}}(\bfk,t) = 0$ and that this results in the introduction of the 
tensor $\MOM{\alpha\beta\gamma}{k}$. Formally, it is called the `momentum transfer operator' (see \cite{McComb90}) and is defined as
\begin{eqnarray}\label{MOM}
\MOM{\alpha\beta\gamma}{k} & \equiv & \frac{1}{2i}\bigg(\nmk_{\beta}\da_{\alpha\gamma} - \frac{\nmk_{\alpha}\nmk_{\beta}\nmk_{\gamma}}{\nmk^2} + \nmk_{\gamma}\da_{\alpha\beta} - \frac{\nmk_{\alpha}\nmk_{\beta}\nmk_{\gamma}}{\nmk^2}\bigg)\label{MOMDef_eqn}\\
&&\nonumber\\
 & \equiv & \frac{1}{2i}\bigg(\nmk_{\beta}P_{\alpha\gamma}(\bfk) + \nmk_{\gamma}P_{\alpha\beta}(\bfk) \bigg).\label{MOMDef2}
\end{eqnarray}
The tensors in the last line are a result of the isotropy and are defined to be 
\beq
\Pj{\al\ba}{k} = \da_{\al\ba}-\frac{\mathrm{k}_{\al}\mathrm{k}_{\ba}}{\mathrm{k}^2} \,.
\eeq
We use this formulation of the NSE to then derive an expression for the spectral correlation function,
\beq
\Ccxc{\al\al'}{k}{t,t'}\delta(\bfk + \bfk') \equiv \crln{\Uuuu{\al\,}{\!}{k}{t}\Uuuu{\al'\,}{\!}{k'}{t'}},
\eeq
which we can later inverse-transform to real-space. However, it is 
well-known that the nonlinear term prevents a statistical treatment of 
the NSE from giving a closed set of equations; an equation for 
the (two-point) correlation 
function requires knowledge of the three-point correlation function, and 
so on. The formalisms described in this paper are attempts to circumvent 
this problem in a systematic way. We first turn to a famous example of a 
Renormalized Perturbation Theory.

\subsection{Example: The Direct Interaction Approximation}\label{dia}
Kraichnan's Direct Interaction Approximation (DIA) 
\cite{Kraichnan57,Kraichnan58,Kraichnan59} pioneered the
renormalized perturbation theory approach to homogeneous 
isotropic turbulence. The basic idea of this approach was to focus
on two quantities, the velocity correlation function and the
response propagator function, and obtain approximate equations to
calculate them.

The main hypothesis of the DIA was that the wave-vector interactions in 
the nonlinear term would be dominated by a single triad 
satisfying $\bfk=\bfj+\bfl$.  Kraichnan exploited this concept and used 
it to create a unique perturbation 
expansion that could be used to bring about a closure to the 
statistical hierarchy. Ultimately, he 
derived a closed set of equations that use only these so-called directly 
interacting wave-vectors.
In the final form, the set includes an equation of motion for the exact 
correlation function of velocity-field coefficients,
\bey\label{DIA_corr}
&&\Big(\dt+\nu\nmk^2\Big)\Ccxc{\al\al'}{k}{t,t'} = \MOM{\al\ba\ga}{k}\!\!\!\!\int_{\bfj+\bfl=\bfk}\!\!\!\!\!d^3\nmj \,d^3\nml\,\times\nonumber\\
&& \qquad\Bigg\{2\!\!\int_{-\infty}^{t}\!\!ds\;\Rrxr{\ba\ba'}{j}{t,s}\MOM{\ba'\da\ea}{j}\Ccxc{\ga\da}{l}{t,s}\Ccxc{\ea\al'}{-k}{s,t'}\nonumber\\
&& \qquad-\int_{-\infty}^{t'}\!\!ds\;\Rrxr{\al'\bar{\al}}{-k}{t',s}\MOM{\bar{\al}\ba'\ga'}{k}\Ccxc{\ga\ga'}{l}{t,s}\Ccxc{\ba\ba'}{j}{t,s}\Bigg\},
\eey
along with an evolution equation for a quantity known as the renormalized propagator function, $\Rrxr{\al\al'}{k}{\!t,t'}$,
\bey\label{DIA_prop}
&&\Big(\dt+\nu\nmk^2\Big)\Rrxr{\al\al'}{k}{\!t,t'} - 4\MOM{\al\ba\ga}{k}\!\!\!\!\int_{\bfj+\bfl=\bfk}\!\!\!\!\!d^3\nmj \,d^3\nml\,\times\nonumber\\
&& \quad\Bigg\{\int_{t'}^{t}\!\!ds\;\Rrxr{\ba\ba'}{j}{t,s}\MOM{\ba'\ga'\bar\al}{j}\Ccxc{\ga\ga'}{l}{t,s}\Rrxr{\bar{\al}\al'}{-k}{s,t'}\Bigg\}\nonumber\\
&& \qquad = \Pj{\al\al'}{k}\da(t-t').
\eey
We give these equations without derivation so that comparisons can be made later in the following sections; more information can be found in his original papers as well as in Beran\cite{Beran68}, McComb\cite{McComb90,McComb95}, Kida and Gotoh \cite{KiGo97}, and Krommes \cite{Krommes02}. Leslie's book \cite{Leslie73} is largely dedicated to Kraichnan's works from this period and provides many insights.

The DIA, although successful in low- to moderate-Reynolds numbers, fails to 
produce a Kolmogorov inertial range. Kraichnan himself showed that the 
DIA gives an inertial range with $\nmk^{-3/2}$ \cite{Kraichnan58} and 
suggested that the DIA does not properly deal with the indirect 
interactions \cite{Kraichnan64,Kraichnan66}, manifest by the DIA's 
inability to decouple the large-scales from the viscous 
scales \cite{Lesieur90, McComb90, McComb95}. As will later be seen the momentum transfer terms 
are in effect vertex functions. The indirect interactions are 
intrinsically associated with these vertex functions leading to the 
notion that ``the whole problem of strong turbulence is contained in a 
proper treatment of vertex renormalization''\cite{MaSiRo73}.

The success and failings of the DIA led to further closures based on renormalized perturbation theories. Notable ones are those of 
Nakano \cite{Nakano72}, 
and the Local Energy Transfer (LET) theory of 
McComb \cite{McComb74,McComb74-2,McComb76,McComb78,McComb84,McComb89,McComb90,McFiSh92,ObMcQu01,McQu03,KiMc04,McKi05}; the latter being the only purely Eulerian theory which is compatible with the Kolmogorov inertial range.
Convinced of the perceived intrinsic failings 
of the DIA based on an Eulerian framework, Kraichnan reformulated fluid 
dynamics to use Lagrangian variables and produced the 
Lagrangian-DIA \cite{Kraichnan65}. This also led to many off-shoots 
notably those of Kraichnan \cite{Leslie73,McComb90,Orszag06} and the 
LRA of Kaneda \cite{Kaneda81,Kaneda86} and Kida and Goto
\cite{KiGo97}. The Eulerian-DIA still 
enjoys some use notably in the regularized-DIA (RDIA) of 
Frederiksen \cite{FrDa00,OKaFr04,FrDan04,FrOKa05}. For completeness, we should mention the separate line of
development, the functional formalism of Hopf \cite{Hopf52},
leading to the self-consistent theories of 
Edwards \cite{Edwards64} and Herring \cite{Herring65,Herring66}.

The sections to follow detail the two formalisms of Wyld and
MSR that aim to achieve a 
statistical closure which properly deals with both the direct and indirect 
interactions. Both formalisms can be applied more generally to
classical dynamics systems,
but in their original papers were applied to turbulence.
It will be argued that both formalisms, although having
different methodology, are equivalent; it will be explicitly shown that
both formalisms
obtain at lowest order the DIA and also agree up to, and
including, fourth order in the perturbative expansion.

\section{The Wyld Formalism}\label{WyldSection}

The Wyld formalism \cite{Wyld61} is a perturbative analysis of the statistical turbulence. It represents one of the earlier attempts \cite{Nakano72,Teodorovich94} to extend the methods of quantum field theory (QFT), specifically those of diagrammatic representation, to the problem of classical turbulence. 

In the approach used by Wyld, the
velocity-field is expanded in a perturbation series, 
with an associated diagrammatic representation,
which then is used in a statistical average to obtain
the two-point velocity correlation function. 
As the 
perturbation series are in fact infinite, a systematic renormalization 
method is employed to reduce the series into a more manageable form. This 
results in integral equations, which at lowest non-trivial order, 
reproduce the Kraichnan DIA result.
We show in subsect. \ref{WyldSubsection2} that Wyld's original renormalization
procedure suffered double-counting issues as noted by Lee \cite{Lee65},
and offer a formal procedure that can remedy this problem.

\subsection{Wyld's Perturbation Method}\label{WyldRedu}
The main focus in examining Wyld's method is the renormalization procedure. 
However, the fundamental or `bare' equations must be established prior 
to renormalization. The following briefly explains Wyld's construction of 
the velocity correlation function via a perturbation expansion of the 
velocity field.
There are a few places in which the following summary of Wyld's 
method deviates from his original work and this
will be pointed out below.  However, these differences
do not change in any essential way his original analysis.

\subsubsection{Wyld's perturbation expansion}
Once again, we start with the NSE in Fourier space,
\begin{equation}\label{Our_WyldPT_2}
\Big(\dt +\nu\nmk^2\Big) u_\al(\mathbf{k},t)=f_\al(\mathbf{k},t)+ \MTOO{\al\ba\ga}{k}\igl{j} u_\ba(\mathbf{j},t) u_\ga(\mathbf{k-j},t).
\end{equation}
{It must be pointed out that our approach already differs somewhat from Wyld's} in that he also
Fourier transformed the time variable. As such, in the Wyld analysis
the wave-vector $\bfk$ and wave-frequency $\om$ are then lumped together 
into a 4-vector $k\equiv(\bfk,-\om)$ and the tensorial NSE is abandoned in 
favor of a simpler one-dimensional `model' equation.  We will not make
any of these changes, and rather work directly with the 3d NSE.
However we will follow the basic formalism set up by Wyld.

Inverting the linear differential operator on the $LHS$ of 
\eqref{Our_WyldPT_2} to the $RHS$ results in the following form of the NSE,
\begin{eqnarray}\label{Our_WyldPT_3}
\Uuuu{\al}{\!}{k}{t} & = & \Rigls{t'}{-\infty}{t}\Rrot{\al\al'}{k}\FFF{\al'}{k}{t'}\nonumber\\
& + & \la\Rigls{t'}{-\infty}{t}\Rrot{\al\al'}{k}\MTOO{\al'\ba\ga}{k}\igl{j}\Uuuu{\ba}{\!}{j}{t'}\Uuuu{\ga}{\!}{k- j}{t'}.\nonumber\\
\end{eqnarray}
A bookkeeping parameter, $\la$, has been multiplied to the nonlinear term 
for the purposes of the perturbation expansion; it will later be set 
equal to unity.
The next step is to consider a perturbation expansion of the NSE,
\begin{equation}\label{Our_Pertub_2_2}
 u_\al(\mathbf{k},t)=\Uuuu{\al}{(0)}{k}{t}+\la\Uuuu{\al}{(1)}{k}{t}+\la^2\Uuuu{\al}{(2)}{k}{t}+\dots,
\end{equation}
This can be substituted for each velocity field in Eq. \eqref{Our_WyldPT_3} and then expressions can be matched by powers of $\la$. At the lowest order, 

\begin{equation}\label{Our_WyldPT_2a}
\Uuuu{\al}{(0)}{k}{t}=\Rigls{t'}{-\infty}{t}\Rrot{\al\al'}{k}\FFF{\al'}{k}{t'},
\end{equation}
and one can establish a response function $\Rror{\al\ba}{k}{t,t'}$ such that 
\begin{equation}\label{Our_Pertub_2_4c}
\Big(\dt +\nu\nmk^2\Big)\Big[\Rror{\al\ba}{k}{t,t'}\Big] = \Pj{\al\ba}{k}\da(t- t').
\end{equation}

Already, it may be seen that there are many variables, arguments, and indices to keep track of, therefore it is useful here to introduce a reduced notation: 
\begin{subequations}
\begin{eqnarray}
\Uuuu{\al}{(0)}{k}{t} & \rightarrow & \uu{k}{0}\\
&&\nonumber\\
\Rror{\al\ba}{k}{t,t'} & \rightarrow & \rro{k}\\
&&\nonumber\\
\MTOO{\al\ba\ga}{k}\igl{j} & \rightarrow & \mmo{k}.
\end{eqnarray}
\end{subequations}
This notation will be less cumbersome for the reader to follow; vector 
indices and time arguments can be determined later where needed. 
The spectral NSE in the new notation becomes
\begin{equation}\label{Our_Pertub_2_1}
(\dt +\nu\nmk^2)\unn{k}=\ff{k}+\la\mmo{k}\unn{j}\unn{k- j},
\end{equation}
{and the equations, Eq. \eqref{Our_WyldPT_3} and 
\eqref{Our_Pertub_2_2}, used for the perturbative treatment are now
respectively,}
\begin{eqnarray}\label{Our_Pertub_2_1}
\unn{k} & = & \rro{k}\ff{k}+\la \rro{k}\mmo{k}\unn{j}\unn{k- j}\\
\unn{k} & = & \uu{k}{0}+\la\uu{k}{1}+\la^2\uu{k}{2}+\la^3\uu{k}{3}\dots
\end{eqnarray}
In using this notation, the integral following the momentum transfer operator 
is always a convolution, where the wave-vector arguments of the convoluted 
functions must add up to the wave-vector of the momentum transfer operator 
immediately preceding them. Some care may be initially needed to keep 
track of these integrated wave-vectors; a simple rule that adjusts for 
this is that all non-$\bfk$ wave-vectors are integrated out.

The perturbation terms by order in $\lambda$ are
\begin{subequations}
\begin{eqnarray}\label{Our_Pertub_2_7}
\lambda^0:  \quad \uu{k}{0} & = & \rro{k} f_{\bfk}\\
\lambda^1:  \quad \uu{k}{1} & = & \lambda \rro{k} \mmo{k}\uu{j}{0}\uu{k- j}{0}\\ 
\lambda^2:  \quad \uu{k}{2} & = & \lambda^2 \rro{k} \mmo{k} \big(2\uu{j}{0}\uu{k- j}{1}\big)\\
&\vdots&\nonumber \,.
\end{eqnarray}
\end{subequations}
The term $\uu{k- j}{1}$ in the integrand of the expression for $\uu{k}{2}$ can be replaced by its definition, leaving $\uu{k}{2}$ written only in terms of $\uuo{\!\;}$. In fact, any order $\uu{\!\;}{n}$ may be written as product of $(n+1)$ $\uuo{\!\;}$'s. For example, the last term in the above equation for the perturbation expansion of $\uu{k}{2}$ can now be written as	
\beq\label{lambda2_ex}
\lambda^2:\;\uu{k}{2}  =  2\lambda^2\rro{k}\mmo{k}\rro{k- j}\mmo{k- j}\uuo{j}\uuo{l}\uuo{k- j- l}.
\eeq

\subsubsection{Wyld's correlation}

Wyld approaches the correlation of two velocity fields very simply by considering the average of the product of two velocity field expansions of $u$'s:
\begin{eqnarray}\label{Our_WyldPT_5}
\Big\langle\Uxuu{\al}{k}{t}\Uxuu{\om}{k'}{t'}\Big\rangle & = & \bigg\langle\Big(\Uouu{\al}{k}{t}+\la\Uuuu{\al}{(1)}{k}{t} + \;\cdots\Big)\times\nonumber\\
&&\Big(\Uouu{\om}{k'}{t'}+\la\Uuuu{\om}{(1)}{k'}{t'} + \;\cdots\Big)\bigg\rangle
\,.
\end{eqnarray}
The zeroth-order and exact correlators are given respectively by
\begin{eqnarray}
\Ccot{\al\om}{k}\delta(\bfk'+\bfk) & \equiv & \crln{\Uuuu{\al}{(0)}{k}{t}\Uuuu{\om}{(0)}{k'}{t'}} = \crln{\uuo{k}\uuo{k'}},\label{Our_WyldPT_8}\\
&& \nonumber\\
\Ccxt{\al\om}{k}\delta(\bfk'+\bfk) & \equiv & \crln{\Uuuu{\al\,}{\!}{k}{t}\Uuuu{\om\,}{\!}{k'}{t'}} = \crln{\unn{k}\unn{k'}},\label{Exact_Correlator_Def}
\end{eqnarray}
and their reduced-notation counterparts are
\begin{eqnarray}
\Ccot{\al\om}{k}\delta_{\mathbf{k}+\mathbf{k'}} & \to & \cco{k}\dl{k}{k'},\\
&& \nonumber\\
\Ccxt{\al\om}{k}\delta_{\mathbf{k}+\mathbf{k'}} & \to & \ccx{k}\dl{k}{k'}.
\end{eqnarray}
The delta-function on the $LHS$ of these definitions is a result of the construction of the Fourier-transform of the real-space correlation equation,
\begin{eqnarray}\label{Our_WyldPT_more}
\crln{\Uxuu{\al}{k}{t}\Uxuu{\om}{k'}{t'}} & = & \frac{1}{(2\pi)^6}\CRLN{\int\!\!d^3\mathrm{x}\int\!\!d^3\mathrm{r}\,\Uxuu{\al}{x}{t}\Uxuu{\om}{x+r}{t'}e^{-i\bfk'\cdot\mathbf{x}}e^{-i\bfk\cdot(\mathbf{x}+\mathbf{r})}}\nonumber\\
& = & \frac{1}{(2\pi)^6}\int\!\!d^3\mathrm{x}\int\!\!d^3\mathrm{r}\,\Crln{\Uxuu{\al}{0}{t}\Uxuu{\om}{r}{t'}}e^{-i(\bfk'+\bfk)\cdot\mathbf{x}}e^{-i\bfk\cdot\mathbf{r}}\nonumber\\
& = & \frac{1}{(2\pi)^3}\int\!\!d^3\mathrm{r}\,\Ccxc{\al\om}{r}{t,t'}e^{-i\bfk\cdot\mathbf{r}}\delta(\bfk'+\bfk)\nonumber\\
& = & \Ccxc{\al\om}{k}{t,t'}\delta(\bfk'+\bfk).
\end{eqnarray}
The second line uses the homogeneity constraint, $\Crln{\Uxuu{\al}{x}{t}\Uxuu{\om}{x+r}{t'}}=\Crln{\Uxuu{\al}{0}{t}\Uxuu{\om}{r}{t'}}$.

Note that the zeroth-order velocity field expansion terms are random Gaussian 
functions as is implied by the temporal delta-function in 
\eqref{Our_Pertub_2_4c}.
Recall the correlation of an odd-numbered product of random Gaussian variables
vanishes. Thus in our analysis this means all terms in the perturbation
expansion with an odd power of $\la$ will vanish.

Continuing in reduced notation (without odd-order moments), a series-expansion for the exact correlator is obtained,
\begin{eqnarray}\label{Our_WyldPT_7}
\crln{\unn{k}\unn{k'}} & = & \cco{k}\dl{k}{k'}\nonumber\\
& + & \la^2\big(\crln{\uuo{k}\uu{k'}{2}} + \crln{\uu{k}{1}\uu{k'}{1}} + \crln{\uu{k}{2}\uuo{k'}}\big) \nonumber\\
& + & \la^4\big(\crln{\uuo{k}\uu{k'}{4}} + \crln{\uu{k}{1}\uu{k'}{3}}+\crln{\uu{k}{2}\uu{k'}{2}} + \crln{\uu{k}{3}\uu{k'}{1}} + \crln{\uu{k}{4}\uuo{k'}}\big)\nonumber\\
& + & \order{6}.
\end{eqnarray}
As mentioned above, all terms can be written as products of zeroth-order 
velocity fields.  For example consider the last of the second-order 
correlations (or moments):
\begin{equation}\label{Our_WyldPT_9}
\crln{\uu{k}{2}\uuo{k'}} = 2\rro{k}\mmo{k}\rro{k-j}\mmo{k-j}\crln{\uuo{j}\uuo{l}\uuo{k-j-l}\uuo{k'}} \,.
\end{equation}
Another property of random-Gaussian variables is that any $n$-th order 
moment may be decomposed into a sum of products of lesser-order moments. 
In the above case, the fourth-order moment is decomposed into three 
pairs of second-order moments,
\begin{eqnarray}\label{Our_WyldPT_13}
\crln{\uu{k}{2}\uuo{k'}} & = & 2\rro{k}\mmo{k}\rro{k-j}\mmo{k-j}\times\nonumber\\
&& \Big(\WCfour{\uuo{j}}{\uuo{l}}{\uuo{k-j-l}}{\uuo{k'}}\Big).\nonumber\\
\end{eqnarray}
Note that all possible combinations of second-order moments are created in 
this decomposition. What is immediately useful here is that the fourth-order 
moment can be written as pairs of second-order moments, which  
more importantly are
zeroth-order correlation functions.

Using the definition of the (zeroth-order) correlator, the above equation becomes 
\begin{equation}\label{Our_WyldPT_142}
\crln{\uu{k}{2}\uuo{k'}}  =   2\rro{k}\mmo{k}\rro{k- j} \mmo{k-j} \Big(\cco{j}\cco{k}\dkk + \cco{j}\cco{k}\dkk + \cco{l}\cco{j}\delta_{\bfk\mms\bfj}\delta_{\bfj+\bfk'}\Big).
\end{equation}
The last term vanishes as it implies $\mmo{k-j}\Bdfnt{k-j}$, which vanishes by definition. Cleaning up leaves
\begin{equation}\label{Our_WyldPT_143}
\crln{\uu{k}{2}\uuo{k'}}   =   4\rro{k}\mmo{k}\rro{k-j}\mmo{k-j}\cco{j}\cco{k}\dkk.
\end{equation}
A similar calculation can be made for the other terms, giving the correlation equation to second-order,
\begin{eqnarray}\label{Our_WyldPT_19o}
\ccx{k}\dl{k}{k'} & = & \cco{k}\dl{k}{k'}\nonumber\\
& + & 4\rro{k}\mmo{k}\rro{k-j}\mmo{k-j}\cco{j}\cco{k}\dkk\nonumber\\
& + & 4\rro{k'}\mmo{k'}\rro{k'-j'}\mmo{k'-j'}\cco{k}\cco{j'}\dkk\nonumber\\
& + & 2\rro{k}\mmo{k}\rro{k'}\mmo{k'}\cco{j}\cco{k-j}\dkk \nonumber\\
& + & \order{4}.
\end{eqnarray}
This can be applied to all orders.  However, as this is an infinite expansion, 
a full calculation will be intractable. This led Wyld to use a diagrammatic 
resummation, an approach that would contain the effects of all orders 
generated by the perturbation expansions into a more manageable set 
of equations.

\subsection{Wyld's Diagrammatic Method}\label{WyldSubsection1}
Wyld associated diagrams to the terms of the perturbation expansion. 
These diagrams could then be combined to form  order-by-order
the velocity correlation.
This procedure in turn produced
a graphical expansion for the exact correlator 
function. In this subsection this diagrammatic formulation is presented
and the resulting expressions for correlation function based on these
diagrams is given.

\subsubsection{Defining diagrams}

Wyld diagrammatic notation assigns particular symbols 
to the various terms present in the perturbation expansion
 of the velocity field: 
\begin{equation}\label{DiagramArray}
\begin{picture}(30,10)
\put(6,3){\includegraphics[scale=1.0]{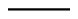}}
\put(13,-1){$\scriptscriptstyle \al$}
\put(11,6){$\scriptstyle \bfk,t$}
\end{picture} \quad\; \leftrightarrow \quad \; \Uout{\al}{k} \quad\quad\; \rightarrow \quad \uuo{k}  \quad \; \begin{array}{l}\text{External}\\ \text{velocity leg}\end{array}\qquad
\end{equation}
\begin{equation}\label{DiagramArray2}
\begin{picture}(30,10)
\put(1,2){\includegraphics[scale=1.0]{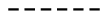}}
\put(0,-2){$\scriptscriptstyle \al$}
\put(27,-3){$\scriptscriptstyle \ba$}
\put(2,4){$\scriptstyle t$}
\put(28,4){$\scriptstyle t'$}
\put(13,5){$\scriptstyle \bfk$}
\end{picture} \quad \; \leftrightarrow \quad \Rrot{\al\ba}{k} \quad \rightarrow \quad \rro{k} \quad \text{Bare propagator}
\end{equation}
\begin{equation}\label{DiagramArray2}
\begin{picture}(30,10)
\put(9,-2){\includegraphics[scale=1.0]{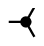}}
\put(6,-2){$\scriptscriptstyle \al$}
\put(12,7){$\scriptscriptstyle \ba$}
\put(12,-5){$\scriptscriptstyle \ga$}
\put(2,2){$\scriptstyle \bfk$}
\put(19,7){$\scriptstyle \bfj$}
\put(19,-7){$\scriptstyle \bfk\mms\bfj$}
\end{picture} \quad \leftrightarrow \quad\; \MTOO{\al\ba\ga}{k} \qquad \rightarrow \quad \mmo{k}\quad \text{Bare vertex} \,.\quad
\end{equation}
These are placed into the relevant equations for the perturbed expressions.

Following the example of Eq. \eqref{lambda2_ex}, the second-order velocity term in the perturbation expansion,
\begin{equation}
\uu{k}{2}=2\rro{k}\mmo{k}\uuo{j}\rro{k\sms j}\mmo{k\sms j}\Big(\uuo{l}\uuo{k\sms j\sms l}\Big),
\end{equation}
can be written in diagrams as 
\begin{equation}\nonumber
\includegraphics[scale=1.0]{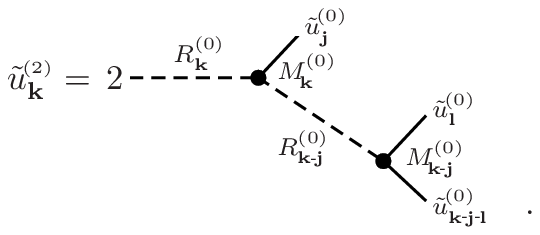}
\end{equation}
The effectiveness of this diagrammatic approach is appreciated by seeing it with all variables, arguments, and indices restored,
\begin{equation}\nonumber
\includegraphics[scale=1.0]{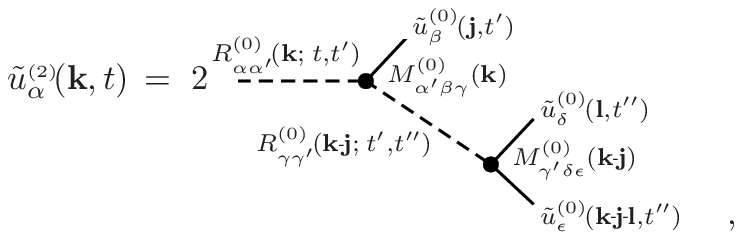}
\end{equation}
for which the corresponding analytic equation is
\begin{eqnarray}
\Uuuu{\al}{(2)}{k}{t} & = & 2 \Rigls{t'}{-\infty}{t}\Rror{\al\al'}{k}{t,t'} \MOM{\al'\ba\ga}{k}\igl{j}\Uouu{\ba}{j}{t'}\times\nonumber\\
&&\qquad \Rigls{t''}{-\infty}{t'}\Rror{\ga\ga'}{k-j}{t',t''}\MTOO{\ga'\da\ea}{k-j}\igl{l}\Big(\Uouu{\da}{l}{t''} \: \Uouu{\ea}{k-j-l}{t''}\Big).
\end{eqnarray}

\subsubsection{`Tree' Diagrams}
Tree diagrams in the perturbation expansion always begin on a propagator 
and end in external velocity legs. The equations and their diagrammatic 
representations for these terms are given below to fourth-order:
\begin{equation}
\includegraphics[scale=1.0]{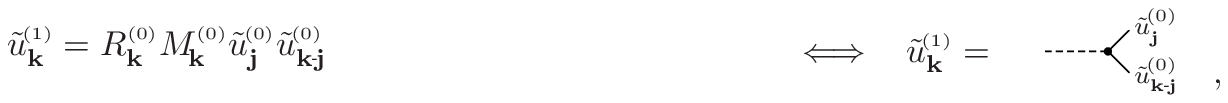}
\end{equation}
\begin{equation}
\includegraphics[scale=1.0]{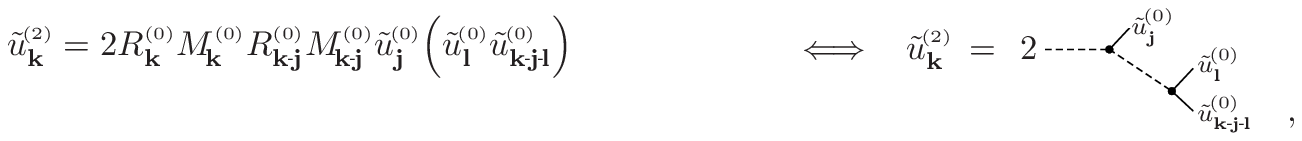}
\end{equation}
\begin{equation}
\includegraphics[scale=1.0]{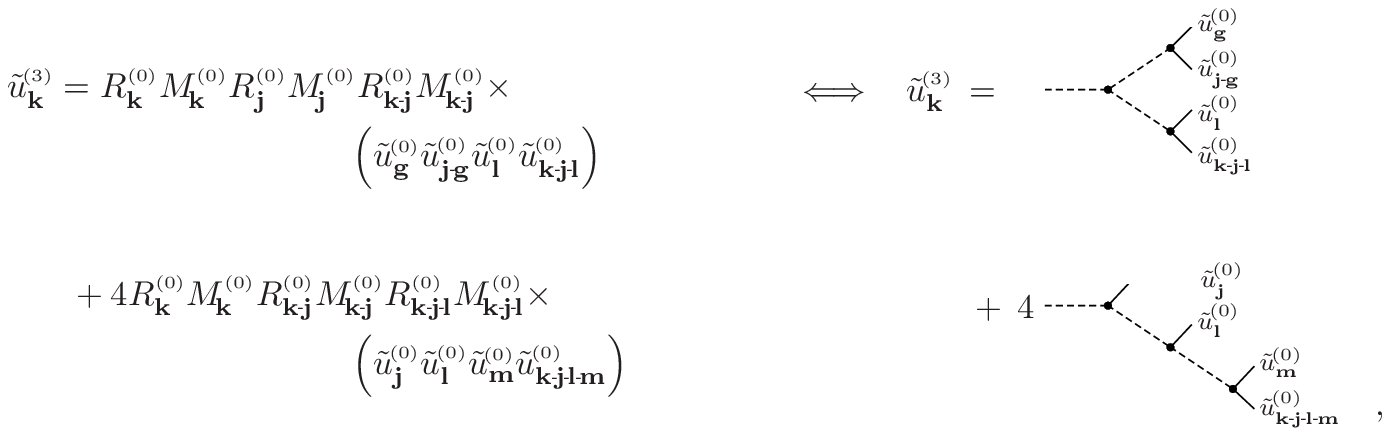}
\end{equation}
\begin{equation}
\includegraphics[scale=1.0]{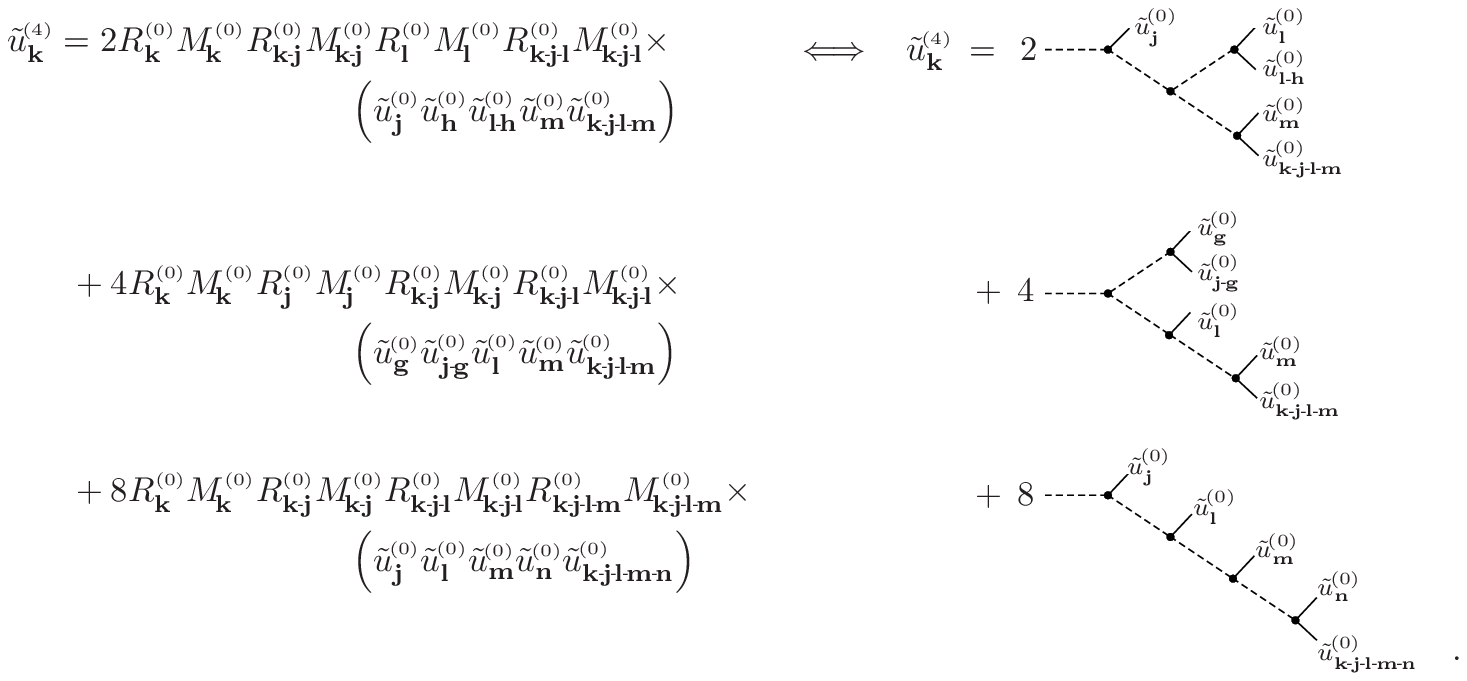}
\end{equation}
Due to their apparent shape, these are the so-called ``tree'' diagrams, with single propagator lines on the left as trunks and propagators ending in zeroth order velocity lines as branches on the right.

\subsubsection{Correlation diagrams} 
Correlation diagrams arise from attaching tree-level diagrams together by fusing the velocity field terms at the ends of branches. These become the zeroth-order correlation functions.
The diagram for the zeroth-order correlation term or `bare correlator' is given by
\begin{equation}
\begin{picture}(205,15)
\put(0,4){\includegraphics[scale=1.0]{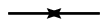}}
\put(45,3){$\leftrightarrow\;\Ccot{\al\ba}{k}$}
\put(-1,1){$\scriptscriptstyle \al$}
\put(26,0){$\scriptscriptstyle \ba$}
\put(1,7){$\scriptstyle t$}
\put(27,7){$\scriptstyle t'$}
\put(12,8){$\scriptstyle \bfk$}
\put(125,3){$ \rightarrow \;\; \cco{k}$.}
\end{picture}
\end{equation}
A similar diagram is used for the exact correlation,
\begin{equation}
\begin{picture}(205,15)
\put(0,3.5){\includegraphics[scale=1.0]{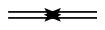}}
\put(45,3){$\leftrightarrow\;\Ccxt{\al\ba}{k}$}
\put(-1,1){$\scriptscriptstyle \al$}
\put(26,0){$\scriptscriptstyle \ba$}
\put(1,9){$\scriptstyle t$}
\put(27,9){$\scriptstyle t'$}
\put(12,10){$\scriptstyle \bfk$}
\put(125,3){$ \rightarrow \;\; \ccx{k}$,}
\end{picture}
\end{equation}
where the double lines are used to distinguish it from its bare counterpart. 

To see how the diagrams operate, it is instructive to examine the 
construction of second-order correlation terms from the tree-level 
diagrams. The first term considered here is the last of the second-order 
terms in Eq. \eqref{Our_WyldPT_7},
\begin{eqnarray}\label{2_9a}
\crln{\uu{k}{2}\uuo{k'}} & = & 2\rro{k}\mmo{k}\rro{k\sms j}\mmo{k\sms j}\crln{\uuo{j}\uuo{l}\uuo{k\sms j\sms l}\uuo{k'}}\nonumber\\
 & = & 4\rro{k}\mmo{k}\rro{k\sms j}\mmo{k\sms j}\cco{j}\cco{k'}\dkk.\quad\;\;
\end{eqnarray}
Diagrammatically, this corresponds to
\begin{equation}
\includegraphics[scale=1.0]{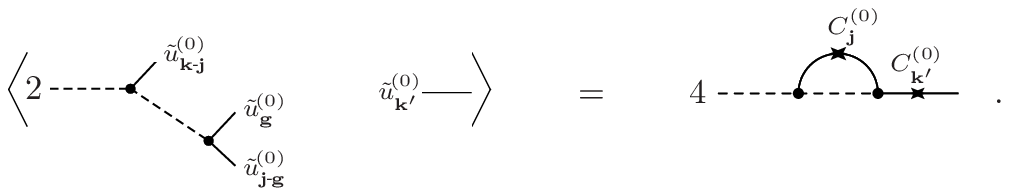}
\end{equation}
Note that we will treat many of the diagrams in this paper as equations. An extra factor of 2 arises from the combinatorics, analogous to the Wick contractions in QFT of the fourth-order moment into products of second-order moments; for example we consider the combinations of the second-order term from above
\begin{equation}
\includegraphics[scale=1.0]{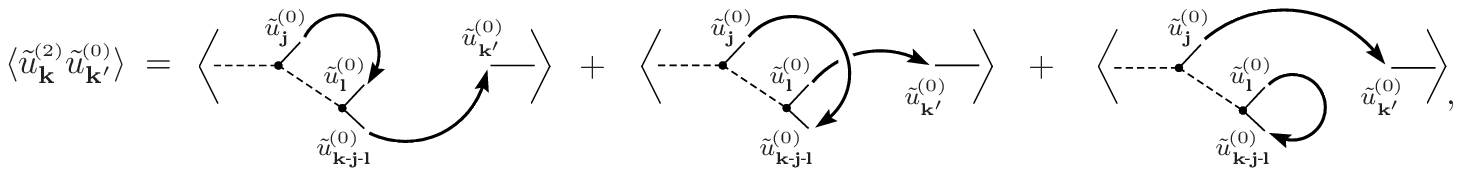}
\end{equation}
which results in 
\begin{equation}
\begin{picture}(392,10)
\put(0,0){$\crln{\uu{k}{2}\uuo{k'}}\;\propto\;$}
\put(65,0){$\crln{\uuo{j}\uuo{l}}\crln{\uuo{k-j-l}\uuo{k'}}$}
\put(166,0){$+$}
\put(194,0){$\crln{\uuo{j}\uuo{k-j-l}}\crln{\uuo{l}\uuo{k'}}$}
\put(296,0){$+$}
\put(324,0){$\crln{\uuo{j}\uuo{k'}}\crln{\uuo{l}\uuo{k-j-l}}$}
\end{picture}
\qquad.
\end{equation}
The last term above vanishes as in Eq. \eqref{Our_WyldPT_142}, introducing a rule that any diagram with a closed loop that is connected to the diagram by a single propagator line will vanish.

\begin{figure}[h]
\includegraphics[scale=1.0]{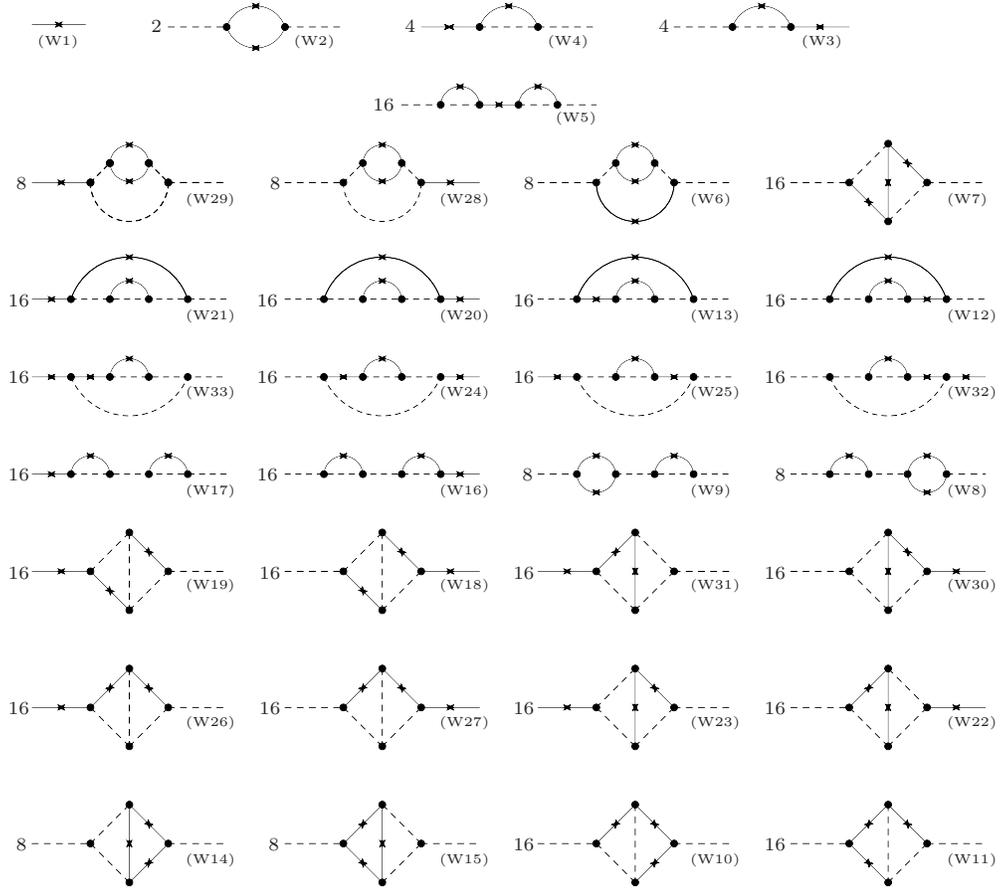}
\caption[Wyld's diagrams representing the correlator expansion.]{Wyld's diagrams representing the correlator expansion up to and including the fourth-order terms. The labels shown on each diagram correspond to those given in Figure 2 of the original paper \cite{Wyld61}.}\label{Correlation_1_3}
\end{figure}

The other diagrams in the second-order terms are obtained by a similar construction,
\begin{eqnarray}\label{2_9b}
\crln{\uuo{k}\uu{k'}{2}} & = & 2\rro{k}\mmo{k}\rro{k'-j'}\mmo{k'-j'}\crln{\uuo{k}\uuo{j'}\uuo{l'}\uuo{k'-j'-l'}}\nonumber\\
 & = & 4\rro{k}\mmo{k}\rro{k'-j'}\mmo{k'-j'}\cco{k}\cco{k'-j'}\dkk\quad\quad\;
\end{eqnarray}
implies
\begin{equation}
\includegraphics[scale=1.0]{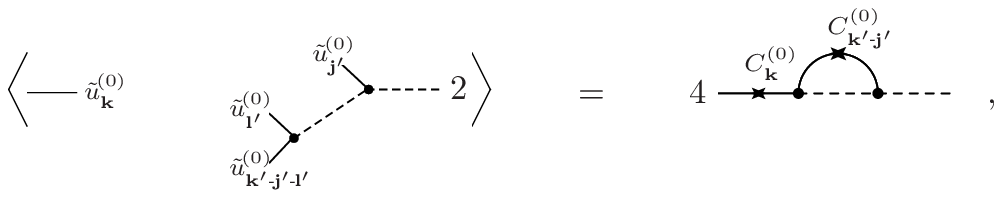}
\end{equation}
and
\begin{eqnarray}\label{2_9c}
\crln{\uu{k}{1}\uu{k'}{1}} & = & \rro{k}\mmo{k}\rro{k'}\mmo{k'}\crln{\uuo{j}\uuo{k\sms j}\uuo{j'}\uuo{k'\sms j'}}\nonumber\\
 & = & 2\rro{k}\mmo{k}\rro{k'}\mmo{k'}\cco{j}\cco{k\sms j}\dkk\quad\;
\end{eqnarray}
implies
\begin{equation}
\includegraphics[scale=1.0]{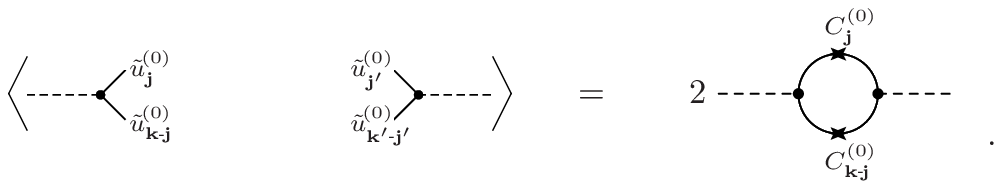}
\end{equation}
This procedure can be applied to all orders, and it can be shown that a 
one-to-one correspondence is established between diagrams and their 
analytical counterparts, with the correct numerical prefactors. A 
reproduction of primitive correlator diagrams to fourth-order is given 
in Fig. \ref{Correlation_1_3}.  

In Wyld's perturbation method all correlations of bare quantities involve Gaussian statistics but at all orders in the perturbation expansions it is formally exact. However this means retaining an infinite number of terms in the expansion of the correlation function. The next subsection sees the systematic renormalization of these terms into a manageable formula.
\newpage
\newpage
\subsection{{Improved Wyld-Lee Renormalized Perturbation Theory}}
\label{WyldSubsection2}
Wyld's method of renormalization is a resummation of diagrams based on the 
emergence and recurrence of fundamental, irreducible diagram units. 
These are so-defined since they are diagrams of a given order 
which can be found as parts of diagrams at all higher orders. To illustrate 
this, we use diagram (W6) in Figure \ref{Correlation_1_3} as an example; 
this diagram is essentially diagram (W2) in which the top correlator has 
been replaced with diagram (W2); the result is a reducible diagram of 
$\order{4}$ composed of an irreducible diagram of $\order{2}$ embedded into 
itself. This will be applied not only to correlators but to the propagator 
and vertex functions as well, introducing additional renormalized equations 
for these quantities. By this process, the resulting system of equations 
is constructed only in terms of renormalized quantities, establishing a 
new renormalized perturbation expansion. An analogy can be made with finding the irreducible diagrams in other diagrammatic methods, most appropriately those in particle physics \cite{PeSc95}.

The main controversies associated with Wyld's formalism are due to the 
renormalization, and therefore, the procedure here will be different 
from Wyld's.  In Wyld's approach for the correlator, he directly
computes the 2-point velocity correlator, inputting the perturbation
expansion at each order for the two velocity fields.  
Instead, we will deduce the two-point velocity correlation function
through the Navier-Stokes equation.
In particular, the new starting point 
{is that which was used by Kraichnan for DIA \cite{Kraichnan57}
(and McComb for the Local Energy Transfer (LET) theory \cite{McComb90}}).
Namely we look at the spectral NSE 
multiplied by a second velocity-field coefficient, 
$u_{\al'}(\mathbf{-k},t')$ and then averaged,
\bey\label{New_C_1}
&&\Big(\PDEop{t}+\nu\nmk^2\Big)\Crln{u_\al(\mathbf{k},t)u_{\al'}(\mathbf{-k},t')} = \Crln{f_\al(\mathbf{k},t)u_{\al'}(\mathbf{-k},t')}\nonumber\\
&& \qquad + \MTOO{\al\ba\ga}{k}\igl{j} \Crln{u_\ba(\mathbf{j},t) u_\ga(\mathbf{k-j},t)u_{\al'}(\mathbf{-k},t')}.
\eey
With the definition of the exact correlator, Eq. \eqref{Exact_Correlator_Def}, 
the above equation can be viewed as an evolution equation for the exact 
correlator, eventually leading to the spectral energy equation. To compute 
the above equation, we apply the perturbation technique to the terms on 
the $RHS$, the force-velocity and the three-point velocity correlation 
functions. This approach was not originally used by Kraichnan but was 
found as an alternate route to deriving the DIA by Leslie \cite{Leslie73}. 

When this correlation expansion is renormalized, there are
two ways it differs from Wyld.
First when the linear operator 
on the $LHS$ is taken to the $RHS$, it remains a bare response function 
(see on the $LHS$ of Eq. \eqref{New_C_1} in its inverted form) and this term 
within the renormalization procedure will always remain unrenormalized. 
Second, the vertex associated with the momentum-transfer operator 
$\MTOO{\al\ba\ga}{k}$ is outside the average, and will also not be included 
in the resummation.  By our above procedure both these points are 
accommodated in a natural way into the formulation.  Part of this
renormalization procedure has been previously pointed out
in the book by McComb \cite{McComb90}, where the first of
the above two steps was noted.
Keeping the vertex function on the $RHS$ of Eq. (\ref{New_C_1})
in its bare form, is being noted here for the first time.
As we will discuss below, both these steps are necessary to
avoid double counting effects in the renormalization procedure.

{The process of renormalization firstly requires the identification of reducible and irreducible diagrams. The first criterion by which to classify diagrams  accordingly} is the ability to 
separate a diagram into two parts by severing a single correlator. 
Diagrams that can be split into two separate diagrams by cutting a 
single correlator are labelled by Wyld as `Class-A'. {We introduce a modification to the procedure here which is to further distinguish} two types of Class-A diagrams: those diagrams 
with the correlators on the $LHS$ of the left-most vertex, and those with 
separable diagrams connected by a single bare correlator that occurs 
on the $RHS$ of the left-most vertex. These are labelled Class-A$_L$ and 
Class-A$_R$, respectively. 

For purposes that will soon become apparent, we note that bare correlators are given analytically by
\begin{equation}\label{Our_WyldPT_8-}
\crln{\Uuuu{\al}{(0)}{k}{t}\Uuuu{\om}{(0)}{k'}{t'}} = \Crln{\Rror{\al\al'}{k}{t,s}\FFF{\al'}{k}{s}\FFF{\om'}{k'}{s'}\Rror{\om\om'}{k'}{t',s'}},
\end{equation}
which reveals the propagators within each. This identification allows the derivation of the propagator diagrams to come. 
An example of cutting an external correlator is given diagrammatically by
\begin{equation}\nonumber
\begin{picture}(250,40)
\put(9.5,25){\includegraphics[scale=1.0]{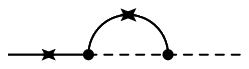}}
\put(20.2,25){\textcolor{Grey}{\pmb{\Big|}}}
\put(97,25){$\to$}
\put(125,25){$\Big($\includegraphics[scale=1.0]{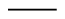}}
\put(145,25){$\Big)$}
\put(144,25){\includegraphics[scale=1.0]{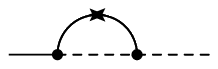}$\Big)$}
\put(148,25){$\Big($}

\put(97,0){$\to$}

\put(124.5,0){$\Big($\includegraphics[scale=1.0]{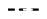}}
\put(145,0){$\Big)$}
\put(144,0){\includegraphics[scale=1.0]{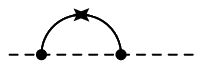}$\Big),$}
\put(148,0){$\Big($}

\put(139.5,1){$\scriptstyle \ff{k}$}
\put(151,1){$\scriptstyle \ff{a}$}
\end{picture}
\end{equation}
with its analytic counterpart given by
\begin{eqnarray}\nonumber
\rro{\sms k}\mmo{\sms k}\rro{j}\mmo{j}\cco{k}\!\!\!\!\!\!\!\!\!\textcolor{Grey}{\pmb{\Big|}}\;\;\;\;\cco{\sms k\sms j}& \to & \Big(u^{\scriptscriptstyle{(0)}}_{\mathbf{k}}\Big)\! \Big(\rro{\sms k}\mmo{\sms k}\rro{j}\mmo{j}\cco{\sms k\sms j}u^{\scriptscriptstyle{(0)}}_{\mathbf{a}}\Big)\nonumber\\
& \to & \Big(\rro{k}\ff{k}\Big) \! \Big(\rro{\sms k}\mmo{\sms k}\rro{j}\mmo{j}\cco{\sms k\sms j}\rro{a}\ff{a}\Big)\nonumber.
\end{eqnarray}
This diagram, (W4), is a member of the Class-A$_L$ diagrams; note that the equation still reads left to right. Using Fig. 2.2 as a reference, the Class-A$_L$ diagrams are: 4, 17-25(odd), 26, 29-33(odd); and the Class-A$_R$ diagrams are: 3, 5, 16-24(even),27, 28-32(even). {While it is not evident, it must be pointed out that the set of Class-A$_L$ diagrams is smaller than its complement but without this classification diagrams will be created redundantly as was found already at fourth-order by Wyld\cite{Wyld61} and Lee \cite{Lee65}. It is for this reason that we have introduced this subdivision of Wyld's Class-A diagrams.}

Note that Class-A$_L$ diagrams can be written with the leftmost correlator in the form of a force-force correlation. We introduce an equation for these diagrams to fourth-order, first graphically followed by its analytic counterpart,
\begin{equation}
\includegraphics[scale=1.0]{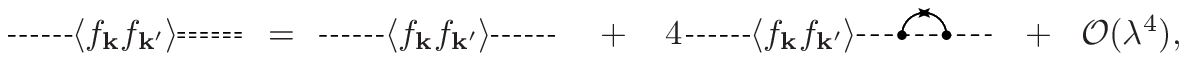}
\end{equation}
\beq
\rro{k}\crln{\ff{k}\ff{k'}}\rrx{k'} \; = \; \rro{k}\crln{\ff{k}\ff{k'}}\rro{k'}\; + \; 4 \rro{k}\crln{\ff{k}\ff{k'}}\rro{k'}\rro{k'}\mmo{k}\rrx{j}\mmo{j}\cco{k'}\;+ \;\order{4}.
\eeq
In writing these equations, a new function, the exact propagator, has been introduced and a new diagram is associated with it,
\begin{equation}
\includegraphics[scale=1.0]{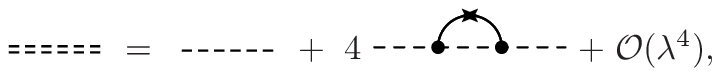}
\end{equation}
\begin{equation}
\rrx{k}\;=\;\rro{k}\;+\;\;4\la^2\rro{k}\mmo{j}\cco{j}\rro{k\sms j}\mmo{k}\rro{k}\;+\;\order{4}.
\end{equation}
The complete expansion of the propagator to fourth-order is given below:
\begin{equation}\label{prop_expansion}
\includegraphics[scale=1.0]{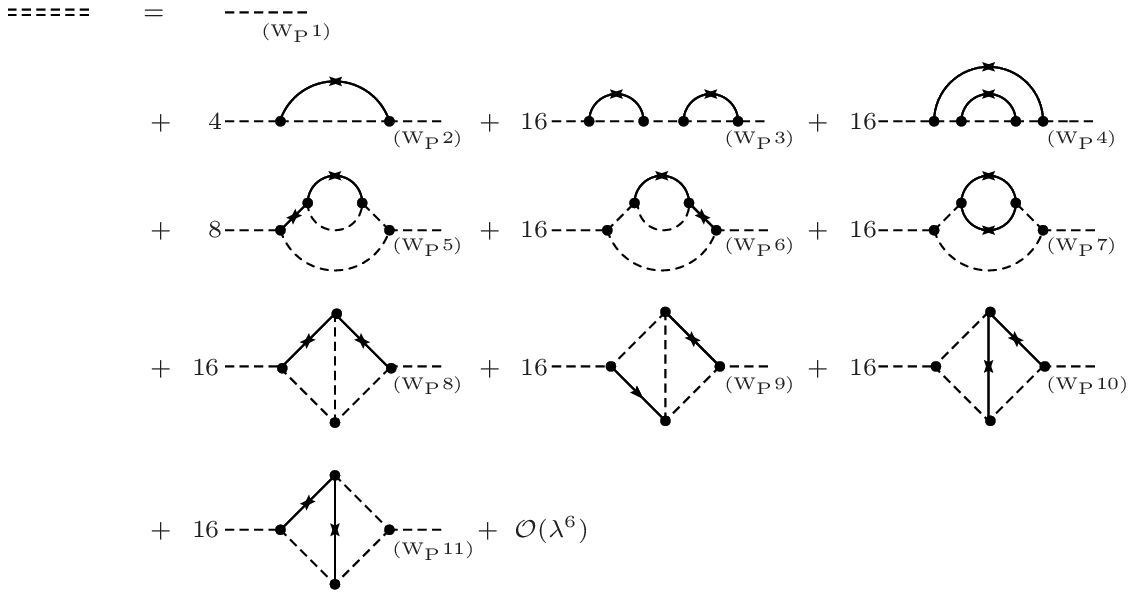}
\end{equation}
{Note that while the method used here to obtain these terms 
is different from Wyld's, in the end the same terms are obtained 
for the expansion. In 
determining this expansion, we have only considered a subset of Wyld's 
Class-A diagrams and enforced that these diagrams have a bare propagator on 
their left, breaking the leftmost correlation function that characterize the 
Class-A$_R$ diagrams. As suggested by McComb \cite{McComb90}, this maintains 
consistency with the DIA derivation of 
Kraichnan \cite{Kraichnan58, Kraichnan59, 
Leslie73}. A similar prescription will be used in determining the renormalized 
equation for the propagator as we will later see.}

The procedure for dealing with Class-A$_R$ diagrams will be postponed as it is similar to that used to re-sum the remaining diagrams. Diagrams not classified as Class-A are designated as Class-B diagrams; these are further classified into reducible and irreducible based on finding embedded elements of low-order within diagrams of higher-order. An example of this can be seen by examining the diagram W6 in Fig.~\ref{Correlation_1_3},
\begin{equation}\nonumber
\includegraphics[scale=1.0]{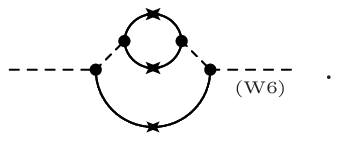}
\end{equation}
It is readily seen that the elements between the two outermost vertices is the diagram W2 in Fig.~\ref{Correlation_1_3}; this is also equivalent to the W2 diagram by replacing the top correlator with itself. This will be given in more detail below however it will be useful to include another function at this time. 

Wyld introduced an exact vertex function as an expansion without giving 
a detailed account of how it was derived but instead wrote a diagram 
series expansion for the exact vertex function,
\begin{equation}\label{exact_vert_func}
\includegraphics[scale=1.0]{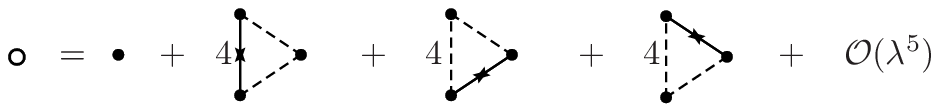}
\end{equation}
\begin{eqnarray}
\mm{k} & = & \mmo{k} \;+\;\; 4\mmo{k}\cco{j}\rro{k\sms j}\mmo{a}\rro{a}\mmo{b}\nonumber\\
& + & 4\mmo{k}\cco{k\sms j}\rro{j}\mmo{a}\rro{a}\mmo{b} \;+\;\; 4\mmo{k}\rro{j}\rro{k\sms j}\mmo{a}\cco{a}\mmo{b}\;+\;\order{5}.
\end{eqnarray}
The renormalized vertex diagram is defined in the same manner as its bare counterpart,
\begin{equation}
\includegraphics[scale=1.0]{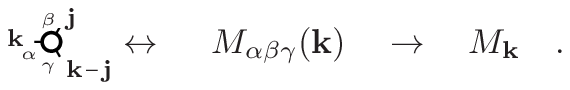}
\end{equation}

Starting with these irreducible diagrams, Wyld generated the full expansion by replacing any of the constituent elements with a higher-order element. For example, replacing the correlator, a propagator, or a vertex in the first term of Eq.\eqref{exact_vert_func} results in the following terms, respectively:
\begin{equation}\nonumber
\begin{picture}(150,40)
\put(40,0){,}
\put(100,0){,}
\put(160,0){.}
\put(0,0){\includegraphics[scale=0.65]{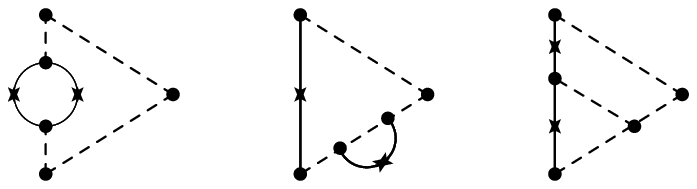}}
\end{picture}
\end{equation}
Irreducible diagrams can then be found by selectively removing correlators, propagators, and vertex corrections from Class-B diagrams. The set is left with diagrams that cannot be constructed from non-trivial (bare correlators, propagators, or vertices) elements. Under this classification, there are two up to fourth order,
\begin{equation}\nonumber
\begin{picture}(190,40)
\put(10,4){\includegraphics[scale=1.05]{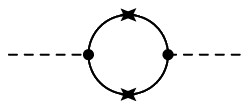}}
\put(65,11){$\scriptstyle (\mathrm{W}2)$}
\put(85,18){$,$}
\put(95,0){\includegraphics[scale=1.0]{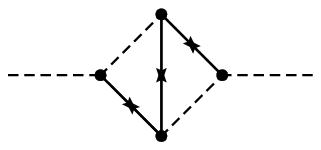}}
\put(168,11){$\scriptstyle (\mathrm{W}7)$}
\put(190,18){.}
\end{picture}
\end{equation}

Starting with the second-order Class-B irreducible diagrams, all Class-B diagrams (except the Class-B irreducible diagrams that arise at higher orders) can be generated by replacing the appropriate correlator, propagator, and vertex corrections with their respective expansions. This is demonstrated using three correlator diagrams,
\begin{equation}\nonumber
\begin{picture}(300,40)
\put(0,1){\includegraphics[scale=0.9]{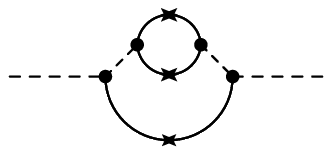}}
\put(69,13){$\scriptscriptstyle (\mathrm{W}6)$}
\put(88,17){$,$}

\put(106,6.6){\includegraphics[scale=0.9]{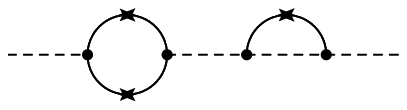}}
\put(195,13){$\scriptscriptstyle (\mathrm{W}9)$}
\put(214,17){$,$}

\put(230,0){\includegraphics[scale=0.9]{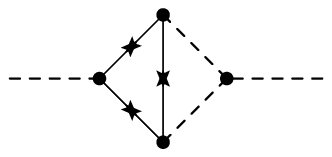}}
\put(295,13){$\scriptscriptstyle (\mathrm{W}15)$}
\put(320,17){$.$}
\end{picture}
\end{equation}

The three correlator diagrams are constructed as follows:

\paragraph{Construction with a Correlator Correction}
\begin{equation}
\begin{picture}(300,40)
\put(10.7,33){$(\quad)$}
\put(0,23){\includegraphics[scale=0.4]{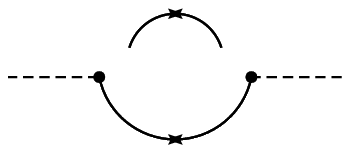}}
\put(45,28){$\mathbf{\rightarrow}$}
\put(77.5,33){$(\quad\:)$}
\put(64,20){\includegraphics[scale=0.5]{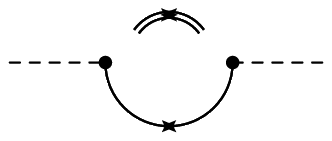}}
\put(119,28){$ = $}
\put(134,24){\includegraphics[scale=0.6]{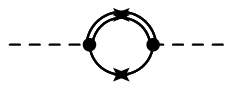}}

\put(119,9.5){$ = $}
\put(137.5,5){\includegraphics[scale=0.6]{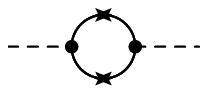}}
\put(186,9.5){$+$}
\put(202,-0.5){\includegraphics[scale=0.6]{./B_R_2a.eps}}
\put(264,9.5){$+\;\;\cdots$}
\put(249,4){$\scriptscriptstyle (\mathrm{W}6)$}
\end{picture}
\end{equation}
Replacing a bare correlator with an exact one in the irreducible second-order correlator diagram is equivalent to inserting the series for the correlator, giving rise to the anticipated diagram as well as others.

\vspace{0.2cm}

\paragraph{Construction with a Propagator Correction}
\begin{equation}
\begin{picture}(340,35)
\put(28,26.1){$(\quad\:\,)$}
\put(30.1,28.5){\includegraphics[scale=0.6]{./R_W_0.eps}}
\put(0,20){\includegraphics[scale=0.6]{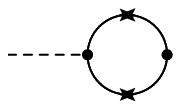}}
\put(60,26.1){$\mathbf{\rightarrow}$}
\put(77.8,20){\includegraphics[scale=0.6]{./B_R_1b.eps}}
\put(105.5,26.1){$(\quad\;\,)$}
\put(108.,28.){\includegraphics[scale=0.6]{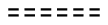}}
\put(138,26.1){$ = $}
\put(154,20){\includegraphics[scale=0.6]{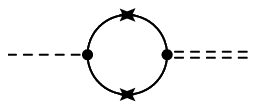}}
\put(138,6){$ = $}
\put(154,0){\includegraphics[scale=0.6]{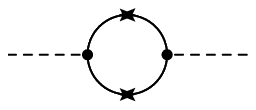}}
\put(206.5,6){$+$}
\put(223,0){\includegraphics[scale=0.6]{./B_R_1a.eps}}
\put(300,6){$+\;\;\cdots$}
\put(278,3){$\scriptscriptstyle (\mathrm{W}9)$}
\end{picture}
\end{equation}
Inserting the exact propagator and its expansion obtains the desired term, (W9).

\vspace{0.2cm}

\paragraph{Construction with a Vertex Correction}
\begin{equation}
\begin{picture}(325,42)
\put(0,22){\includegraphics[scale=0.6]{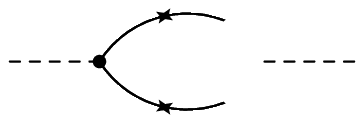}}
\put(34.5,31){$(\;)$}
\put(37.6,32.2){\includegraphics[scale=0.6]{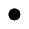}}
\put(70,31.2){$\mathbf{\rightarrow}$}
\put(90,22){\includegraphics[scale=0.6]{./B_R_3b.eps}}
\put(124,31){$(\;)$}
\put(126.4,31.3){\includegraphics[scale=0.8]{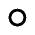}}
\put(160,31){$ = $}
\put(178,24.9){\includegraphics[scale=0.6]{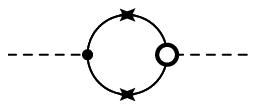}}

\put(160,10){$ = $}
\put(178,4){\includegraphics[scale=0.6]{./C_W_2b2.eps}}
\put(228,10){$+$}
\put(245,0){\includegraphics[scale=0.6]{./B_R_3a_2.eps}}
\put(310,10){$+ \;\;\, \cdots$}
\put(284,8){$\scriptscriptstyle (\mathrm{W}15)$}
\end{picture}
\end{equation}
As with the others, the inclusion of the vertex expansion in this case produces (W15) plus others at higher orders.

In the above examples, the leftmost propagator and vertex have remained unrenormalized. This procedure can be applied to the Class-B and Class-A$_R$ diagrams in Fig.~\ref{Correlation_1_3}, generating the respective irreducible diagrams. Replacing the appropriate bare diagrams with the renormalized quantities in the irreducible diagrams and then collecting these together with the Class-A$_L$ terms expressed in Eq. \eqref{prop_expansion}, leads to an equation for the exact correlator,
\begin{equation}\label{LeeKW_corr_full}
\includegraphics[scale=1.0]{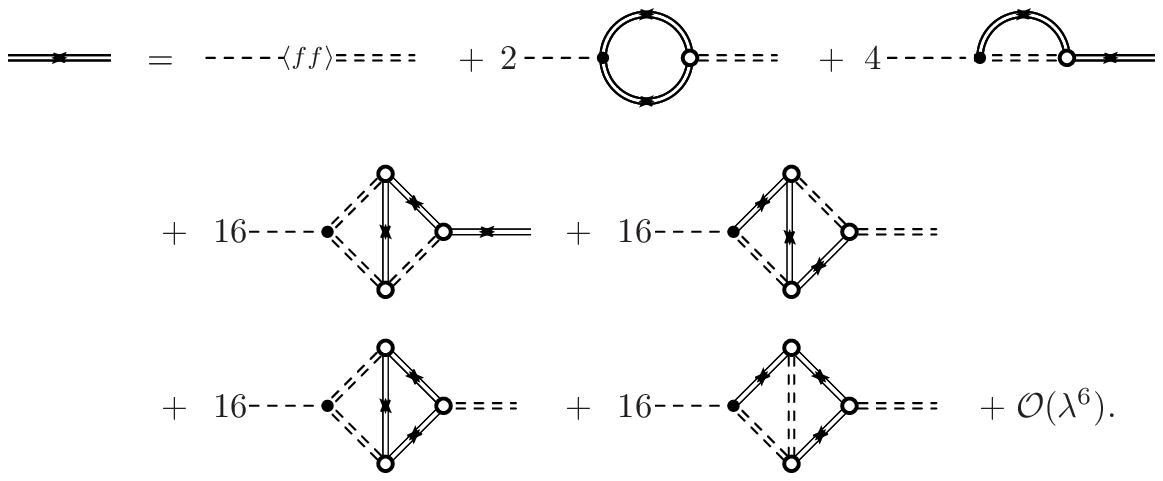}
\end{equation}
We note the distinction between Wyld's original exact correlator equation 
(Fig. 5 in \cite{Wyld61}),
\begin{equation}\label{w-3_ii}
\begin{picture}(400,70)
\put(0,38.5){\includegraphics[scale=1.0]{./CC_W.eps}}
\put(36,39){$\displaystyle =$}
\put(55,40){\includegraphics[scale=1.0]{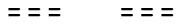}}
\put(72,39.6){$\scriptstyle \crln{ff}$}
\put(114,39){$+\;\;2$}
\put(134,24.5){\includegraphics[scale=1.0]{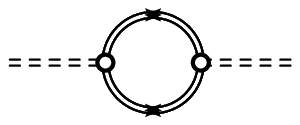}}
\put(221,39){$+\;16$}
\put(242,19){\includegraphics[scale=1.0]{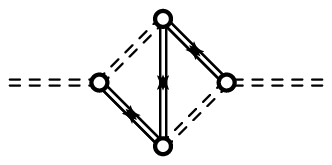}}
\put(340,39){$+\;\order{6}$,}
\put(48,22){$\displaystyle \underbrace{\phantom{p\qquad\quad\quad p}}_{\text{Class-A diagrams}}$}
\put(131,22){$\displaystyle \underbrace{\phantom{p\qquad\qquad\qquad\qquad\qquad\qquad\qquad\qquad\qquad\qquad\qquad\qquad p}}_{\text{Class-B diagrams}}$}
\end{picture}
\end{equation}
where the first term on the $RHS$ contains both A$_L$ and A$_R$ diagrams. 
It can easily be checked that both Eqs. \eqref{LeeKW_corr_full} 
and \eqref{w-3_ii} lead to the same perturbation expansion.
Further remarks about the exact correlator will be made in a later section; 
we now continue with the propagator.

The approach Wyld used to determine an equation for the exact correlator 
could not be used for the exact propagator. Using both Classes A$_L$ and 
A$_R$, the result is a primitive expansion for the propagator with some 
of the terms redundantly generated, specifically diagrams 
($\mathrm{W}_\mathrm{P}$3), ($\mathrm{W}_\mathrm{P}$6), and 
($\mathrm{W}_\mathrm{P}$7) in Eq. \eqref{prop_expansion}. Wyld recognized this problem and circumvented 
this by using a Dyson equation for the propagator, expressing his 
arguments only mathematically and without a diagram equation for the 
exact propagator. The result was to introduce modified vertex functions 
and use the Ward-Takashi identities to relate these to the propagators, 
as in QFT \cite{PeSc95}. Wyld's final complete set of equations has a 
diagrammatic expansion for the exact correlator, two diagram series for 
the exact and modified vertex functions, and the Ward-Takahashi 
identities. 

However, it was argued by Lee that the method of using the Dyson equation and 
Ward-Takahashi identities cannot be applied to the full 3-dimensional NSE 
in a manner similar to Wyld's scalar model \cite{Lee65}. Lee had adapted Wyld's 
method to magnetohydrodynamic turbulence and found the same problem 
but introduced the following equation for the exact propagator:
\begin{equation}
\begin{picture}(250,20)
\put(0,1.5){\includegraphics[scale=1.0]{./RR_W.eps}}
\put(35,0){$=$}
\put(50,2.5){\includegraphics[scale=1.0]{./R_W_0.eps}}
\put(86,0){$+\;\,4$}
\put(104,-1){\includegraphics[scale=1.0]{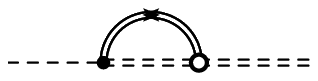}}
\put(198,0){$\;+\;\;\,\order{4}$}
\end{picture}
\end{equation}
The left-most propagator and vertex have both been left bare. The former 
clears the redundant generation of propagator diagram 
($\mathrm{W}_\mathrm{P}3$), while the latter does the same for 
diagrams ($\mathrm{W}_\mathrm{P}6$) and ($\mathrm{W}_\mathrm{P}7$). 
This equation for the propagator does correct the redundancies, however, 
the asymmetry introduced by Lee at second-order also does not generate
 Wyld's diagram (W10) at fourth-order,
hence his inclusion of a fourth-order irreducible propagator term not found 
in Wyld. Lee included this term in the exact propagator,
\begin{equation}\label{LeeKW_prop}
\begin{picture}(360,40)
\put(0,0){\includegraphics[scale=1.9]{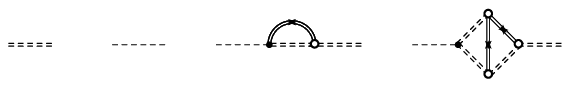}}
\put(38,16.5){$=$}
\put(95,16.5){$+\;\;4$}
\put(201.5,16.5){$+\;16$}
\put(310,16.5){$\;+\;\order{6}$.}
\end{picture}
\end{equation}
Lee introduced this as an {\it ad hoc} fix, however we observe here that
the above equation can be derived by adopting a scheme where the leftmost 
propagator and vertex of the exact propagator remain unrenormalized, 
similar to what we did for the correlator expansion. We have now 
a single procedure for deriving both the
correlator and propagator expansions.

This was the motivation for using Eq. \eqref{New_C_1} at the start of this 
Subsection as our formulation of Wyld's approach. Finding the irreducible 
propagator diagrams from the set of Class-A$_L$ diagrams while 
maintaining that the left-most propagator and vertex remain unrenormalized 
results in Eq. \eqref{LeeKW_prop}. Our justification for doing so follows 
one of Kraichnan's early derivations of the propagator \cite{Kraichnan64}. His approach was to 
include an additional, infinitesimal perturbation, $\eta_\alpha(\bfk,t)$, 
to the equation for the velocity field (c.f. Eq. \eqref{Our_WyldPT_3}),
\begin{eqnarray}\label{Spec_NSE_with_pert}
\Uuuu{\al}{\!}{k}{t} & = & \Rigls{t'}{-\infty}{t}\Rrot{\al\al'}{k}\FFF{\al'}{k}{t'} + \Rigls{t'}{-\infty}{t}\Rrot{\al\al'}{k}\eta_{\alpha'}(\bfk,t')\nonumber\\
& + & \la\Rigls{t'}{-\infty}{t}\Rrot{\al\al'}{k}\MTOO{\al'\ba\ga}{k}\igl{j}\Uuuu{\ba}{\!}{j}{t'}\Uuuu{\ga}{\!}{k- j}{t'},
\end{eqnarray}
and then to derive the propagator as a response function to the perturbation, 
\begin{eqnarray}
\widehat{R}_{\al\om}(\bfk;t,t')\delta(\bfk-\bfk')\;\; & \equiv\ & \;\; \frac{\delta\Uuuu{\al}{\!}{k}{t}}{\delta\eta_{\omega}(\bfk',t')}
\nonumber \\
 & = & \Rrot{\al\om}{k}\delta(\bfk-\bfk')
 +  \la\Rrot{\al\al'}{k}\MTOO{\al'\ba\ga}{k}\igl{j}\frac{\delta}{\delta\eta_{\omega}(\bfk',t')}\bigg(\Uuuu{\ba}{\!}{j}{t'}\Uuuu{\ga}{\!}{k- j}{t'}\bigg).\label{Spec_NSE_with_pert_prop1}\nonumber\\
\end{eqnarray}
This response function would then be averaged in the same manner as the correlation function to give the renormalized response function or propagator, 
\begin{equation}
R_{\al\om}(\bfk;t,t')\;\; = \;\;\Big\langle{\widehat{R}_{\al\om}(\bfk;t,t')}\Big\rangle.
\end{equation}
Note that the averaging would leave the bare propagator and the vertex unchanged,
\begin{equation}\label{Spec_NSE_with_pert_prop2}
R_{\al\om}(\bfk;t,t')\delta(\bfk-\bfk')=\Rrot{\al\om}{k}\delta(\bfk-\bfk') + \la\Rror{\al\al'}{k}{t,s}\MTOO{\al'\ba\ga}{k}\igl{j}\CRLN{\frac{\delta}{\delta\eta_{\omega}(\bfk',t')}\bigg(\Uuuu{\ba}{\!}{j}{s}\Uuuu{\ga}{\!}{k- j}{s}\bigg)};\\
\end{equation}
this analytic form of the propagator is consistent with our current version.
In particular observe that Eq. \eqref{Spec_NSE_with_pert_prop2}
is consistent with our procedure of keeping the leftmost propagator
and vertex functions as bare.\newline

Lastly, we consider the equation for the renormalized vertex as a function of renormalized quantities. Since there is no leftmost vertex or propagator, the exact vertex expansion from Wyld is still valid to $\order{5}$ and can be used here,
\begin{equation}\label{W_vertex_expansion}
\begin{picture}(250,40)
\put(0,11){\includegraphics[scale=1.0]{./GG_0.eps}}
\put(18,11){$=$}
\put(36,11){\includegraphics[scale=1.0]{./G_0.eps}}
\put(55,11){$+\;\;4$}
\put(74,-3){\includegraphics[scale=0.75]{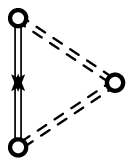}}
\put(105,11){$+\;\;4$}
\put(124,-3){\includegraphics[scale=0.75]{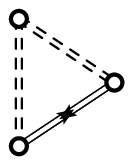}}
\put(155,11){$+\;\;4$}
\put(174,-3){\includegraphics[scale=0.75]{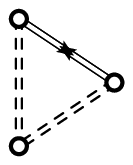}}
\put(205,11){$+\;\;\order{5}$ .}
\end{picture}
\end{equation}

In summary, the exact correlator, given by Eq. \eqref{LeeKW_corr_full} along
with the renormalized propagator and vertex functions, Eqs. \eqref{LeeKW_prop} and \eqref{W_vertex_expansion}, 
respectively, constitute a closed set of integral equations that describe the time evolution of the 
two-point correlation function. We will refer to this set as the
{``Improved Wyld-Lee Renormalized Perturbation Theory''.} It contains the same information about the 
correlator expansion up to fourth-order, Fig.~\ref{Correlation_1_3}, and 
describes turbulence insofar as the NSE can be treated perturbatively, 
with the Gaussian statistics assumed in the external forcing introduced 
to facilitate the closure. The above equations are still, in principle, 
infinite series, however the resummation contains the detail of a greater 
number of terms and allows a truncation that retains more of this information.
Truncating after second-order and noting that the bare propagator 
is equivalent to $(\dt+\nu\nmk^2)^{-1}$, the equation above and the equation 
for the propagator Eq. \eqref{LeeKW_prop} give the result of the DIA, 
Eqs. \eqref{DIA_corr} and \eqref{DIA_prop}. 

Thus we have developed a single unified procedure for deriving
both the correlator and propagator expansions in the Wyld formalism.
The correlator equation from Wyld's original approach, Eq. \eqref{w-3_ii}, is correct, but his approach to calculating
the propagator was flawed, as it led to
the redundant generation of diagrams as observed by Lee \cite{Lee65}.
Our procedure of keeping the leftmost propagator and vertex as bare
recovers Wyld's original and correct correlator expansion
but at the same time leads to a correct equation for the
propagator expansion.  This is our simple single procedure for
obtaining the Wyld theory.

As such, we have established here a well-defined procedure for renormalization. 
Summarizing this procedure 
in terms of the equation for the correlation, we symbolically write it as,
\begin{equation}
{\cal L}_{\scriptscriptstyle{0}}  \langle \tilde{u}\tilde{u}\rangle = M^{\scriptscriptstyle{(0)}} \langle \tilde{u}\tilde{u}\tilde{u}\rangle .
\end{equation}
Note that the linear operator, whose inverse is the bare propagator, and the vertex function are
outside of the correlations. In perturbation theory the 3-velocity correlation function above
would be a function of the bare correlation,
vertex function and response function,
$\langle \tilde{u}\tilde{u}\tilde{u}\rangle = f[C^{\scriptscriptstyle{(0)}},M^{\scriptscriptstyle{(0)}},R^{\scriptscriptstyle{(0)}}]$ and  our renormalization procedure is
to replace these by the exact respective functions
$f[C^{\scriptscriptstyle{(0)}},M^{\scriptscriptstyle{(0)}},R^{\scriptscriptstyle{(0)}}] \rightarrow f[C,M,R]$.
In other words our procedure keeps
the left-most propagator and vertex out of the resummation. 

\section{The Martin-Siggia-Rose Formalism}\label{MSRSection}

The MSR Formalism \cite{MaSiRo73} is by now a well-known analytic procedure 
that can be 
used to calculate the ``statistical dynamics of classical systems''. The 
formalism establishes an operator theory where the observables are defined as 
Heisenberg operators. This permits a non-perturbative treatment akin to the 
Schwinger formalism \cite{Schwinger51a,Schwinger51b,Schwinger51c} in quantum 
field theory, which formally closes the statistical moment hierarchy. The 
operator formalism introduces an adjoint operator, which can be used in the 
construction of nontrivial commutation relations leading to correlation and 
response functions \cite{Krommes02}. Employing the Schwinger formalism for 
statistical closure involves the use of a generating or characteristic 
functional. An alternative to the construction of such operators 
comes {from path integrals \cite{Jensen81, Thacker97}}.

Notable sources providing detailed information on MSR are works by Rose 
\cite{Rose74}, Phythian \cite{Phythian75,Phythian76,Phythian77}, 
Andersen \cite{Andersen00} and Krommes \cite{Krommes02}. As demonstrated 
in their original paper, the formalism is applicable to the turbulence 
problem which has inspired further work in the analysis of least-action 
principles \cite{Eyink96a,Eyink96b} and gauge 
symmetries \cite{BeHo05,BeHo07,BeHo09} in the study of turbulence.

In the previous part of the paper we reformulated Wyld-Lee, in order to
eliminate the double-counting problem in a more natural way. Then we
presented a rather pedagogic clarification of their theory, particularly of
the way in which the use of diagrams led to renormalization. Now, in this
section, we aim to make a comparable elucidation of MSR's theory. Of course
in their case there is no procedural error to rectify. Yet, a pedagogic
exposition of their use of diagrams in obtaining renormalization is of value
and, when taken in conjunction with our exposition of Improved Wyld-Lee
theory, allows the relationship between the two formalisms to be clearly
seen. In this way we are able to show that the two formalisms are fully
equivalent.

\subsection{Setting up the Formalism}
It has been seen already that the velocity field $u_{\al}(\mathbf{x},t)$ is a fundamental observable in fluid dynamics. The MSR formalism extends the common notion of it to that of a classical statistical operator \cite{Phythian75,Andersen00}. In the language of quantum field theory (QFT), it is similar to a Heisenberg operator, in that it is time dependent \cite{PeSc95}. This is an essential first step in establishing the formalism.
\subsubsection{Dynamical Equations}
A generalized equation of motion for a generalized field variable, $\psi(1) \equiv u_{\al_1}(\mathbf{x}_1,t_1)$, is introduced,
\begin{equation}\nonumber
\partial_{t_1}\psi(1) = U(1) + U(12)\psi(2) + U(123)\psi(2)\psi(3).
\end{equation}
It is argued that this equation can accommodate many dynamical systems, and in principle, can be generalized to higher orders of interaction. We are free with the choice of variables and so to maintain consistency with the previous part of this paper, we continue to work in spectral space,
\begin{equation}\label{MSR_GenEOM}
\dt\Uxuu{\al}{k}{t} = \uotu{\al}{k}{t}+\uotu{\al\ba}{k,j}{t,t'}\Uxuu{\ba}{j}{t'} + \uotu{\al\ba\ga}{k,j,l}{t,t',t''}\Uxuu{\ba}{j}{t'}\Uxuu{\ga}{l}{t''}.
\end{equation}
The quantities defined within,
\begin{eqnarray}
\uotu{\al}{k}{t} & \quad \leftrightarrow \quad & \text{0-point potential/external force}\\
\uotu{\al\ba}{k,j}{t,t'} & \quad \leftrightarrow \quad & \text{1-point potential}\\
\uotu{\al\ba\ga}{k,j,l}{t,t',t''} & \quad \leftrightarrow \quad &\text{2-point potential},
\end{eqnarray}
are the generalized interaction potential functions, which may depend on time and be random. Integration of repeated arguments and summation of indices is implied. Using $\delta$-functions for wave-vector- and time-arguments, and Kronecker-$\delta$s where needed, we can specify these potentials to give the spectral NSE: 
\begin{eqnarray}
\uotu{\al}{k}{t} & = & \FFF{\al}{k}{t}\nonumber\\
\Rightarrow F_{\al}(\bfk, t) \,, &&\\
\uotu{\al\ba}{k,j}{t,t'}\Uxuu{\al}{j}{t} & = & -\nu\igl{j}\!\!\int\!\!dt'\delta_{\al\ba}\delta(\bfk-\bfj)\delta(t-t')\bfj^2 \Uxuu{\al}{j}{t}\nonumber\\
\Rightarrow D_{\al\ba}(\bfk,\bfj;t,t') & \equiv & -\nu\igl{j}\!\!\int\!\!dt'\delta_{\al\ba}\delta(\bfk-\bfj)\delta(t-t')\bfj^2 \,, \\
\uotu{\al\ba\ga}{k,j,l}{t,t',t''}\Uxuu{\ba}{j}{t'}\Uxuu{\ga}{l}{t''} & = & \lambda\MTOO{\al\ba\ga}{k}\igl{j}\!\!\igl{l}\!\!\!\int\!\!dt'\!\!\!\int\!\!dt''\times\nonumber\\
& & \qquad \delta(\bfk-\bfj-\bfl)\delta(t-t')\delta(t-t'')\Uxuu{\ba}{j}{t'}\Uxuu{\ga}{l}{t''} \,. \nonumber\\
\Rightarrow M^{\scriptscriptstyle{(0)}}_{\al\ba\ga}(\bfk,\bfj,\bfl;t,t',t'') & \equiv & \lambda\MTOO{\al\ba\ga}{k}\igl{j}\!\!\!\igl{l}\!\!\!\int\!\!dt'\!\!\!\int\!\!dt''\delta(\bfk-\bfj-\bfl)\delta(t-t')\delta(t-t'')
\,.
\end{eqnarray}
We see that the 0-point potential takes the role of the external force, which is uncoupled to the velocity field; the 1-point potential comprises the linear terms of the equation, namely the dissipative term; and the 2-point potential now becomes the convection term. We have re-labelled the potentials to help in the comparison.

In keeping the goal of an analogous formalism to QFT, an adjoint operator is introduced by way of a canonical commutation relation,
\begin{equation}\label{commutator}
\big[\Uxuu{\al}{k}{t},\UDJ{\ba}{k'}{t}\big]=\delta_{\al\ba}\,\delta(\mathbf{k}-\mathbf{k'}),
\end{equation}
thus defining the operator as a functional derivative,
\begin{equation}\label{hatu}
\UDJ{\al}{k}{t}\;\equiv\;\frac{-\delta}{\delta\Uxuu{\al}{k}{t}}.
\end{equation}
In the path-integral formalism, the Fourier conjugate of the adjoint field occurs naturally in the treatment of the delta-functional (see for example Jensen \cite{Jensen81} and Krommes \cite{Krommes02}).
An equation of motion for the adjoint field,
\begin{equation}\label{MSR_EQ_Adj}
-\dt\UDJ{\al}{k}{t} = D_{\ba\al}(\bfk,\bfj;t,t')\UDJ{\ba}{j}{t'} + M^{\scriptscriptstyle{(0)}}_{\ba\ga\al}(\bfj,\bfl,\bfk;t',t'',t)\UDJ{\ba}{j}{t'}\Uxuu{\ga}{l}{t''},
\end{equation}
may be constructed using Eq. \eqref{MSR_GenEOM}, Eq. \eqref{commutator} and the 
canonical relations,
\begin{equation}
\dt\Uxuu{\al}{k}{t}  = \frac{\delta\mathcal{H}[\mathbf{u},\widehat{\mathbf{\tilde{u}}},t]}{\delta\UDJ{\al}{k}{t}}, \qquad\quad -\dt\UDJ{\al}{k}{t} = \frac{\delta\mathcal{H}[\mathbf{u},\widehat{\mathbf{\tilde{u}}},t]}{\delta\Uxuu{\al}{k}{t}};
\end{equation}
the generalized Hamiltonian functional, $\mathcal{H}[\mathbf{u},\widehat{\mathbf{\tilde{u}}},t]$, is postulated and then determined from the first of these relations, see for example \cite{Rose74,Phythian75,Deker79,Eyink96b}.

We can construct a single dynamical equation that includes Eq. \eqref{MSR_GenEOM} and Eq. \eqref{MSR_EQ_Adj}. Both fields are collected together in what Eyink calls a `doublet field'\cite{Eyink96b} and Krommes calls an `extended field vector'\cite{Krommes02}, 
\begin{equation}
\pmb{\mathcal{U}}(\al,\bfk;t)=\left[\! \begin{array}{c} \Uxuu{\al}{k}{t} \\ \UDJ{\al}{k}{t} \end{array}\!\right].
\end{equation}
For these `doublet' quantities, a curly script is used and their doublet-vector indices are given in \textsc{smallcaps} font where the indices are either $+$ or $-$; for example $\spinu{A=+}{\al}{k}{t} = \Uxuu{\al}{k}{t}$. We can also establish a commutator for the doublet field vector,
\begin{equation}
\Big[\spinu{A}{\al}{k}{t},\spinu{B}{\ba}{k'}{t}\Big]=i\paulitwo{AB}\delta_{\al\ba}\,\delta(\mathbf{k}-\mathbf{k'}),
\end{equation}
where the (Pauli) matrix is
\begin{equation}
\pauli\equiv\left[ \begin{array}{rr} 0 & \mms i \\ i & 0 \end{array} \right].
\end{equation}
As argument labels will soon increase, we continue with our reduced notation where all arguments are combined into subscript wave-vectors,  $\spinu{A}{\al}{k}{t}\to\spinnu{k}{t}$; this is effectively the same as the notation used in MSR. 

An equation of motion for the doublet field vector is then simply constructed from the dynamical equations of $\mathbf{u}$ and $\widehat{\mathbf{u}}$,
\begin{equation}\label{spineom}
- i\pauli\dt\spinnu{k}{t} \; = \; \Fotu{k}{t}+\Dotu{k,j}{t,t'}\;\spinnu{j}{t'} +  \frac{1}{2}\Totu{k,j,l}{t,t',t''}\;\spinnu{j}{t'}\;\spinnu{l}{t''}.
\end{equation}
Once again, the curly-script used for the so-called `doublet-field' potentials distinguishes them from their earlier counterparts. The above equation combines the system of equations, Eq. \eqref{MSR_GenEOM} and Eq. \eqref{MSR_EQ_Adj}, into a single equation for a single field variable; it is to this equation that we apply our statistics.

\subsubsection{Statistics}
A generating functional is introduced and used to create all statistical quantities,
\begin{eqnarray}
\mathcal{Z} & = & \timo{\Bigg}{\exp\Big[\int^{t_f}_{t_i} \spinu{A}{\al}{k}{t} \eta_{\scriptscriptstyle\text{\textsc{A}}}(\al,\mathbf{k};t) dt\big]},
\end{eqnarray}
where $\timo{\!\,}{\cdots}$ denotes time-ordering. Here 
$\eta_{\scriptscriptstyle\text{\textsc{A}}}(\al,\mathbf{k},t)$ plays the role of the source term 
that is standard to these techniques.  It is, in effect, a perturbation 
to the 0-point potential (c.f. Eq. \eqref{Spec_NSE_with_pert}). 

Using the generating functional, one can find the statistical moments 
or cumulants as needed through functional differentiation of the generating 
functional with respect to the source term. In practice, the cumulants 
are obtained by functionally differentiating the logarithm of the generating 
functional, returning what are called the connected Green's 
functions \cite{PeSc95}. As an example, the first- and second-order cumulants 
are produced respectively via
\begin{eqnarray}\label{1rstOrdCumu}
G_{\scriptscriptstyle\text{\textsc{A}}}(\alpha;\mathbf{k};t) & = & \frac{\delta}{\delta\eta_{\scriptscriptstyle\text{\textsc{A}}}(\alpha,\mathbf{k},t)}\ln \crln{\mathcal{Z}}\nonumber\\
 & = & \frac{\Crln{\timo{\big}{\mathcal{Z}\mathcal{U}_{\scriptscriptstyle\text{\textsc{A}}}(\alpha,\mathbf{k},t)}}}{\crln{\mathcal{Z}}},
\end{eqnarray}
\begin{eqnarray}\label{2ndOrdCumu}
G_{\scriptscriptstyle\text{\textsc{AB}}}(\alpha,\beta;\mathbf{k,k'};t,t') & = &\frac{\delta^2}{\delta\eta_{\scriptscriptstyle\text{\textsc{A}}}(\alpha,\mathbf{k},t)\delta\eta_{\scriptscriptstyle\text{\textsc{B}}}(\beta,\mathbf{k'},t')}\ln \crln{\mathcal{Z}}\nonumber\\
& = &\frac{\delta}{\delta\eta_{\scriptscriptstyle\text{\textsc{A}}}(\alpha,\mathbf{k},t)}G_{\scriptscriptstyle\text{\textsc{B}}}(\beta;\mathbf{k'};t').
\end{eqnarray}
In the reduced notation, $G_{\scriptscriptstyle\text{\textsc{A}}}(\alpha;\mathbf{k};t)\to\bigG{k}{t}$ and $G_{\scriptscriptstyle\text{\textsc{AB}}}(\alpha,\beta;\mathbf{k,k'};t,t')\to \bigG{k,k'}{t,t'}$.
Note that the correlator and propagator functions are contained within the second-order cumulant of the extended field vector,
\begin{eqnarray}
\pmb{G}(\al,\ba, \mathbf{k,k'};t,t')\Big|_{\eta=0} & = & \left[\begin{array}{cc}
\Crln{\Uxuu{\al}{k}{t}\Uxuu{\ba}{k'}{t'}} & \Crln{\Uxuu{\al}{k}{t}\UDJ{\ba}{k'}{t'}}\\
\Crln{\UDJ{\al}{k}{t}\Uxuu{\ba}{k'}{t'}} & 0
\end{array}\right]\nonumber\\
&& \nonumber\\
& = & \left[\begin{array}{cc}
\Ccxc{\al\ba}{k,k'}{t,t'} & \Rrxr{\al\ba\,}{k,k'}{t,t'}\\
\Rrxr{\ba\al\,}{k',k}{t',t} & 0
\end{array}\right],
\end{eqnarray}
where we recall that $\Uxuu{\al}{k}{t}$ is defined in (\ref{hatu})

The interest is in obtaining a dynamical equation for a particular statistical quantity, which for example can be the second-order correlation function of two velocity fields of a turbulent fluid. Using Eq. \eqref{spineom}, Eq. \eqref{1rstOrdCumu}, and Eq. \eqref{2ndOrdCumu}, one can construct an equation of motion for  the mean field,
\begin{equation}\label{mean}
-i\pauli\dt\bigG{k}{t} = \Fotu{k}{t}+\neta{k}{t}+\Dotu{k,j}{t,t'}\bigG{j}{t'} +\frac{1}{2}\Totu{k,j,l}{t,t',t''}\Big(\bigG{j,l}{t',t''}+\bigG{j}{t'}\bigG{l}{t''}\Big).
\end{equation}
A second-order cumulant is present on the $RHS$ in this equation
for the first-order cumulant on the $LHS$, thus generating a statistical
hierarchy and so the closure problem. 
Differentiating Eq. \eqref{mean} by $\neta{k'}{t'}$ gives
\begin{equation}\label{cumulant2}
-i\pauli\dt\bigG{k,k'}{t,t'} = \delta_{\mathbf{k,k'}}+\Dotu{k,j}{t,t''}\bigG{j,k'}{t'',t'}+\Totu{k,j,l}{t,t'',s}\Big(\bigG{j}{t''}\bigG{l,k'}{s} + \frac{1}{2}\frac{\delta\bigG{j,l}{s,t''}}{\delta\neta{k'}{s'}}\Big).
\end{equation}
In this case, the problem of closure occurs with the last term where a third-order cumulant is introduced,
\begin{equation}
\frac{\delta\bigG{k,j}{s,t'}}{\delta\neta{l}{t''}}=\bigG{k,j,l}{s,t',t''}.
\end{equation}
A method is needed to proceed further without the introduction of {\it ad hoc} hypotheses to link various moments or cumulants.

\subsubsection{Closure}
The problem of closure can now be addressed. The method employed by MSR
is the Schwinger-Dyson formalism, which has been used to deal with the 
closure problem in QFT \cite{PeSc95}.
The authors sought a way to transfer this quantum statistical
formalism to classical physics. By way of a functional Legendre transform,
\begin{equation}
\mathcal{L}\big[G_{\scriptscriptstyle\text{\textsc{A}}}(\alpha;\mathbf{k};t)\big]=\ln\mathcal{Z}\big[\eta_{\scriptscriptstyle\text{\textsc{A}}}(\alpha;\mathbf{k};t)\big]-G_{\scriptscriptstyle\text{\textsc{A}}'}(\alpha';\mathbf{k'};t')\eta_{\scriptscriptstyle\text{\textsc{A}}'}(\alpha';\mathbf{k'};t')
\end{equation}
a closure can be found through the introduction of vertex functions which can be related to cumulants of order-$n$ through $n$-order functional derivatives of the above equation with respect to $\bigG{j}{s}$ \cite{Krommes02}. As we wish to find a closure for the third-order cumulant, it is straightforward to calculate the three-point vertex function,
\begin{eqnarray}\label{vertex1}
\mathcal{M}_{\text{\tiny{ABC}}}(\alpha,\beta,\gamma;\mathbf{k,j,l};t,t',t'') & = & \frac{\delta^3 \mathcal{L}[G_{\text{\tiny{A}}'}(\alpha';\mathbf{k'};s')]}{\delta G_{\text{\tiny{A}}}(\alpha;\mathbf{k};t)\delta G_{\text{\tiny{B}}}(\beta;\bfj;t')\delta G_{\text{\tiny{C}}}(\gamma;\bfl;t'')}\\
&&\nonumber\\
& = & -\frac{\delta}{\delta G_{\text{\tiny{A}}}(\alpha,\mathbf{k};t)}\Big[G_{\text{\tiny{BC}}}(\beta,\gamma;\mathbf{j,l};t',t'')\Big]^{-1}\label{vertex},
\end{eqnarray}
which is symmetric with respect to the indices $\textsc{a}$, $\textsc{b}$ and $\textsc{c}$.
We also note this is different from usual QFT notation,
in that we use $\mathcal{M}$ to represent this function instead of the more 
common $\Gamma$, which is used in both Wyld and MSR; the reason for doing 
this is for its connection to the renormalized vertex function introduced 
earlier. Similarly, the last term in
\eqref{cumulant2} may be rewritten to contain the three-point vertex
\begin{eqnarray}\label{vertexcumulant}
\frac{1}{2}\Totu{k,j,l}{t,t'',s}\frac{\delta\bigG{j,l}{s,t'}}{\delta\neta{k'}{t''}} & = &\frac{1}{2}\Totu{k,j,l}{t,t'',s}\bigG{j,j'}{t'',s''}\bigG{l,l'}{s,s'}\bigg(\frac{-\delta\Ggib{j',l'}{s'',s'}}{\delta\bigG{m}{r}}\bigg)\bigG{m,k'}{r,t'}\nonumber\\
& = &\frac{1}{2}\Totu{k,j,l}{t,t'',s}\bigG{j,j'}{t'',s''}\bigG{l,l'}{s,s'}\Gramma{m,j',l'}{t,t',t''}\bigG{m,k'}{r,t'}.
\end{eqnarray}

Another function that is crucial in QFT is the self-energy 
function \cite{Schwinger48}. In particle physics
self-energy is responsible for attributing 
a particle with a `dressed' or renormalized mass, which is an observable 
quantity. In the present calculation, the self-energy is defined 
using the three-point vertex,
\begin{eqnarray}\label{BigSig_full}
&&\Sigma_{\scriptscriptstyle\text{\textsc{A}}\text{\textsc{A}}'}(\alpha,\alpha';\mathbf{k,k'};t,t')\nonumber\\
&&\quad\equiv\frac{1}{2}\mathcal{M}^{\scriptscriptstyle{(0)}}_{\scriptscriptstyle\text{\textsc{A}}\text{\textsc{B}}\text{\textsc{C}}}(\alpha,\beta,\gamma;\mathbf{k,j,l};t,r,s) G_{\scriptscriptstyle\text{\textsc{B}}\text{\textsc{B}}'}(\beta,\beta';\mathbf{j},\mathbf{j'};r,r') G_{\scriptscriptstyle\text{\textsc{C}}\text{\textsc{C}}'}(\gamma,\gamma';\mathbf{l},\mathbf{l'};s,s')\mathcal{M}_{\scriptscriptstyle\text{\textsc{A}}'\text{\textsc{B}}'\text{\textsc{C}}'}(\alpha',\beta',\gamma';\mathbf{k',j',l'};t',r',s'),
\end{eqnarray}
or in reduced notation, 
\begin{equation}\label{BigSig}
\bigma{k,k'}{t,t'} \equiv \frac{1}{2}\Totu{k,j,l}{t,s,r}\bigG{j,j'}{s,s'}\bigG{l,l'}{r,r'}\Gramma{k',j',l'}{t',s',r'}.
\end{equation}
The dynamical equation for the second-order cumulant, Eq. \eqref{cumulant2}, 
can be rewritten with the self-energy term as,
\begin{eqnarray}\label{cumulantSE}
-i\pauli\dt\bigG{k,k'}{t,t'} & = & \delta_{\mathbf{k,k'}}+\Dotu{k,j}{t,t''}\bigG{j,k'}{t'',t'}+\Totu{k,j,l}{t,t'',s}\bigG{j}{t''}\bigG{l,k'}{s}+\bigma{k,j}{t,s'}\bigG{j,k'}{s',t'}.
\end{eqnarray}
The inclusion of the self-energy leads to the establishment of the well-known Dyson equation \cite{Dyson49a}
\begin{equation}\label{OriginalG}
\Ggibo{k,k'}{t,t'}\bigG{j,k'}{s',t'}=\delta_{\mathbf{k,k'}}+\bigma{k,j}{t,t'}\bigG{j,k'}{s',t'}.
\end{equation}
The Dyson equation is an equation of motion for the second-order cumulant, which is directly obtained from Eq. \eqref{BigSig}. It also establishes the inverse bare second-order cumulant, defined by
\begin{equation}\label{InvG0}
\Ggibo{k,k'}{t,t'}\equiv-i\pauli\partial_{t}\delta_{\mathbf{k,k'}}-\Dotu{k,k'}{t,t'}-\Totu{k,j,k'}{t,t',t''}\bigG{j}{t''}.
\end{equation}
Written differently,
\begin{equation}\label{InvG}
\Ggib{k,k'}{t,t'}=\Ggibo{k,k'}{t,t'}-\bigma{k,k'}{t,t'},
\end{equation}
this equation gives similar (Dyson) equations relating the bare and exact propagators,
\begin{equation}\label{InvGR}
\Rgib{k,k'}{t,t'}=\Rgibo{k,k'}{t,t'}-\big[\bigma{k,k'}{t,t'}\big]_{\pm\mp},
\end{equation}
which are the off-diagonal components in $\big[\bigG{k,k'}{t,t'}\big]^{\mms 1}$;  this equation can be re-arranged to give
\begin{equation}\label{InvGR0}
\bigR{k,k'}{t,t'}=\bigRo{k,k'}{t,t'}+\bigRo{k,j}{t,t'}\big[\bigma{j,j'}{t,t'}\big]_{\pm\mp}\bigR{j',k'}{t,t'}. 
\end{equation}
One can see that the renormalization is already done in the above by the self-energy term. Furthermore, this is the propagator (to second order) established earlier,
\begin{eqnarray}\nonumber
\Rrxr{\al\ba}{k}{t,t'} & = & \Rror{\al\ba}{k}{t,t'}\nonumber\\
& + & \Rror{\al\al'}{k}{t,t''}\bigg(\MTO{\al'\da\ga}{k}\Rrxr{\da\da'}{k-j}{t'',s}\Ccxc{\ga\ga'}{j}{t'',s}\MTO{\da'\ba'\ga'}{-k}\bigg)\Rrxr{\ba'\ba}{-k}{s,t'}. 
\end{eqnarray}
Returning to the vertex function, by differentiating Eq. \eqref{InvG} and using Eq. \eqref{vertex} it can be rewritten in terms of the self-energy and itself:
\begin{equation}\label{BigGam}
\Gramma{k,j,l}{t,t',t''} \; = \; \Totu{k,j,l}{t,t',t''}+\bigg(\frac{\delta\bigma{k,j}{t,t'}}{\delta\bigG{k',j'}{s,s'}}\bigg)\bigG{k',k''}{s,r}\bigG{j',j''}{s',r'}\Gramma{k'',j'',l}{r,r',t''}.
\end{equation}\newline

{This completes our brief exposition of the MSR formalism. We now have a set of equations that close the statistical hierarchy:
\begin{eqnarray}
-i\pauli\dt\bigG{k}{t}& = &\Fotu{k}{t}+\Dotu{k,y}{t,t'}\bigG{j}{t'} +\frac{1}{2}\Totu{k,j,l}{t,t',t''}\Big(\bigG{j,z}{t',t''}+\bigG{j}{t'}\bigG{z}{t''}\Big),\\
-i\pauli\dt\bigG{k,k'}{t,t'} & = & \delta_{\mathbf{k,k'}}+\Dotu{k,j}{t,t''}\bigG{j,k'}{t'',t'}+\Totu{k,j,l}{t,t'',s}\bigG{j}{t''}\bigG{l,k'}{s}+\bigma{k,j}{t,s'}\bigG{j,k'}{s',t'},\\
\bigma{k,k'}{t,t'} & = & \frac{1}{2}\Totu{k,j,l}{t,s,r}\bigG{j,j'}{s,s'}\bigG{l,l'}{r,r'}\Gramma{k',j',l'}{t',s',r'}\\
\Gramma{k,j,l}{t,t',t''} & = & \Totu{k,j,l}{t,t',t''}+\bigg(\frac{\delta\bigma{k,j}{t,t'}}{\delta\bigG{k',j'}{s,s'}}\bigg)\bigG{k',k''}{s,r}\bigG{j',j''}{s',r'}\Gramma{k'',j'',l}{r,r',t''}.
\end{eqnarray}
Going one step further to incorporate the self-energy term,
\begin{eqnarray}
-i\pauli\dt\bigG{k}{t}& = &\Fotu{k}{t}+\Dotu{k,y}{t,t'}\bigG{j}{t'} +\frac{1}{2}\Totu{k,j,l}{t,t',t''}\Big(\bigG{j,z}{t',t''}+\bigG{j}{t'}\bigG{z}{t''}\Big),\\
-i\pauli\dt\bigG{k,k'}{t,t'} & = & \delta_{\mathbf{k,k'}}+\Dotu{k,j}{t,t''}\bigG{j,k'}{t'',t'}+\Totu{k,j,l}{t,t'',s}\bigG{j}{t''}\bigG{l,k'}{s}+\bigg(
\frac{1}{2}\Totu{k,m,n}{t,s,r}\bigG{m,m'}{s,s'}\bigG{n,n'}{r,r'}\Gramma{j,m',n'}{t',s',r'}
\bigg)\bigG{j,k'}{s',t'},\\
\Gramma{k,j,l}{t,t',t''} & = & \Totu{k,j,l}{t,t',t''}+\bigg(\frac{\delta}{\delta\bigG{k',j'}{s,s'}}\Big(
\frac{1}{2}\Totu{k,m,n}{t,s,r}\bigG{m,m'}{s,s'}\bigG{n,n'}{r,r'}\Gramma{j,m',n'}{t',s',r'}
\Big)\bigg)\bigG{k',k''}{s,r}\bigG{j',j''}{s',r'}\Gramma{k'',j'',l}{r,r',t''},
\end{eqnarray}
we find that these equations require iterations which can also extend to infinite orders. Truncations of $\Gramma{k,j,l}{t,t',t''}$ are then necessary to deliver practical and computable results. The first-order truncation of the vertex function is 
\begin{equation}
\Gramma{k,j,l}{t,t',t''} \approx \Totu{k,j,l}{t,t',t''},
\end{equation}
and while it may not be immediately evident, this truncation gives the DIA. The next order of approximation,
\begin{equation}\label{3rdOrderApproxGamma}
\Gramma{k,j,l}{t,t',t''} \approx \Totu{k,j,l}{t,t',t''}+\Totu{k,k',j''}{t,t',t''}\bigG{j'',k''}{s,r}\Totu{j,j',k''}{r,r',t''}\bigG{j',j''}{s,r}\bigG{k',k''}{s',r'}\Totu{k'',j'',l}{r,r',t''},
\end{equation}
similarly generates fourth-order terms given in Wyld, as will also be seen in the following section.}
\subsection{The Diagrammatic Representation of MSR}\label{MSRSubsection1}

The formalism was originally demonstrated with diagrams related to
the turbulence problem, and is by the author's account comparable to that 
of Wyld. We now work directly from their diagrammatic interpretation and 
make a direct comparison to Wyld.


The correlator and propagators are obtained from the second-order cumulant tensor,
\begin{eqnarray}\label{GtoCRR}
\pmb{G}(\alpha,\beta;\mathbf{k,k'};t,t')\Big|_{\eta=0} & = & \left[\begin{array}{cc}
\Ccxc{\al\ba}{k,k'}{t,t'} & \Rrxr{\al\ba\,}{k,k'}{t,t'}\\
\Rrxr{\ba\al\,}{k',k}{t',t} & 0
\end{array}\right].
\end{eqnarray}
We define a diagram for this object just as with the correlator and other diagrams,
\begin{equation}
\begin{picture}(205,16)
\put(0,3.5){\includegraphics[scale=1.1]{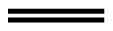}}
\put(45,3){$\leftrightarrow\;\;G_{\text{\tiny{\textsc{AB}}}}(\alpha,\beta; \mathbf{k,k'};t,t')$}
\put(-5.,-1){\tiny{\textsc{A}},$\scriptstyle \alpha$}
\put(25.,-1){\tiny{\textsc{B}},$\scriptstyle \beta$}
\put(-5,11){$\scriptstyle \bfk,t$}
\put(25,11){$\scriptstyle \bfk',t'$}
\put(155,3){$ \rightarrow \;\; \bigG{k,k'}{t,t'}$.}
\end{picture}
\end{equation}
Using the same notation for exact correlators and propagators from the 
Wyld analysis, Eq. \eqref{GtoCRR} can be transcribed diagrammatically as 
\begin{equation}\label{MSR_1C}
\begin{picture}(200,50)
\put(0,10){\includegraphics[scale=1.1]{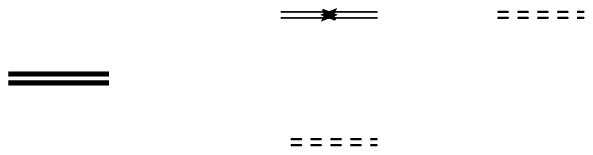}}
\put(54,28.5){$=\;\;\,\,\left[\begin{array}{c} \\ \\ \\ \\ \\ \end{array}\right.$}
\put(168,9){0}
\put(180,28.5){$\left.\begin{array}{c} \\ \\ \\ \\ \\ \end{array}\right]$\;.}
\end{picture}
\end{equation}
The external force introduced as $\Fotu{}{}$ would not survive in the second-order cumulant equation, Eq. \eqref{cumulant2}. This term is needed for stationary turbulence, as was seen in Wyld. MSR avoids this difficulty by introducing the forcing as potential. In the case of turbulence considered here, MSR considered the mean field to be zero, hence $\bigG{k}{}=0$. Noting these points,  
Eq. \eqref{InvG0} is rewritten as
\begin{equation}\label{InvG0_inK}
\Ggibo{k,k'}{t,t'} = -i\pauli\partial_{t}\delta_{\mathbf{k,k'}}-\Dotu{k,k'}{t,t'}-\mathcal{F}_{\mathbf{k,k'}}.
\end{equation}
For the forcing potential $\mathcal{F}_{\mathbf{k,k'}}$ the only 
nonzero component is
$\Big[\mathcal{F}_{\mathbf{k,k'}}\Big]_{--} = \crln{\FFF{\al}{k}{t}\FFF{\al'}{k'}{t'}}$.

Using Eq. \eqref{InvG}, an equation for $\bigG{k,k'}{t,t'}$ can be constructed,
\begin{equation}\label{BigG}
\bigG{k,k'}{t,t'}=\bigG{k,j}{t,t'}\Ggibo{j,j'}{t,t'}\bigG{j',k'}{t,t'}-\bigG{k,j}{t,t'}\bigma{j,j'}{t,t'}\bigG{j',k'}{t,t'}.
\end{equation}
From the structure of these matrices, using the Dyson equation for the propagator given in Eq. \eqref{InvGR0} along with $\bigR{k,a}{t,t'}\Rgib{a,k'}{t,t'}=\delta_{\mathbf{k,k'}}$ \cite{Krommes02}, one can construct a graphical interpretation for the second-order cumulant equation,

\begin{equation}\label{MSR_2C}
\includegraphics[scale=1.0]{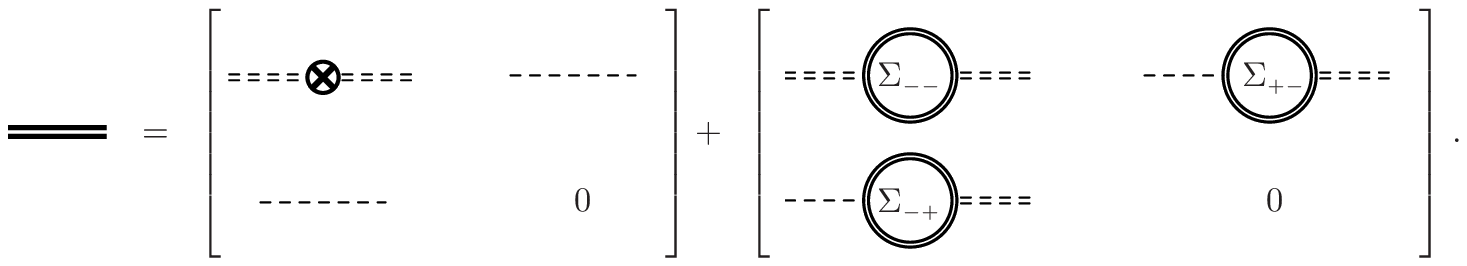}
\end{equation}
Equating Eq. \eqref{MSR_1C} and Eq. \eqref{MSR_2C}, one can obtain a diagrammatic equation for the exact correlator
\begin{equation}\label{MSR_1C_2}
\begin{picture}(280,32)
\put(2,0){\includegraphics[scale=1.0]{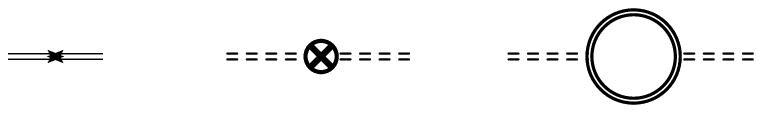}}
\put(175,12){$\Sigma_{\scriptscriptstyle{--}}$}
\put(44,12.5){=\qquad\qquad\quad\qquad\;\; + \qquad\qquad\quad\qquad\;\;\;\;,}
\end{picture}
\end{equation}
where $\Sigma_{\scriptscriptstyle{--}}\;=\;\big[\bigma{k,k'}{t,t'}\big]_{\mms \mms}$ and 
\begin{equation}\label{MSR_1X}
\begin{picture}(180,10)
\put(0,1){\includegraphics[scale=1.0]{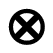}}
\put(28,5){$ = \quad \big\langle\pmb{f}(\mathbf{k},t)\otimes\pmb{f}(\mathbf{k'},t')\big\rangle$}
\end{picture}
\end{equation}
is the external force contribution, which in the case of NSE turbulence is the correlation of two random forces. The mathematical formula for this is then
\begin{equation}\label{Correla}
C_{\mathbf{k,k'}}=R_{\mathbf{k,j}}\big\langle\mathbf{f(j},t)\otimes\mathbf{f(j'},t')\big\rangle R_{\mathbf{j',k'}} + R_{\mathbf{k,j}}\big[\bigma{j,j'}{t,t'}\big]_{\mms \mms}R_{\mathbf{j',k'}}.
\end{equation}

The graphical equation for the exact propagator can likewise be extracted from Eq. \eqref{MSR_2C},
\begin{equation}\label{MSR_1R}
\begin{picture}(195,32)
\put(2,0){\includegraphics[scale=1.0]{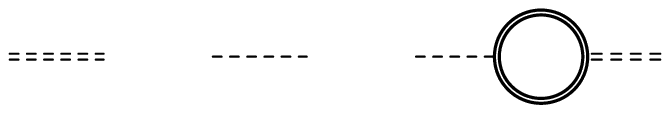}}
\put(148,11.8){$\Sigma_{\scriptscriptstyle{\pm\mp}}$}
\put(44,12.3){=\qquad\qquad\quad  + \qquad\qquad\quad\qquad\;\;\;\;,}
\end{picture}
\end{equation}
with its analytic counterpart Eq. \eqref{InvGR0}.
Note at this point that the MSR procedure has built into it that the 
leftmost response function in Eq. \eqref{MSR_1R} is unrenormalized and 
from Eq. \eqref{BigSig} and Eq. \eqref{MSR_1R} that the leftmost vertex 
function is bare. Recall in the Wyld formulation Lee had introduced 
these corrections \emph{ad hoc}, and we showed in Eq. \eqref{New_C_1} 
how this can be properly accounted for 
at the outset. With this modification to the Wyld procedure, the Wyld 
and MSR renormalized perturbation approaches are equivalent. We will 
show explicitly that this is true up to fourth-order in the 
perturbation expansion. 

We now express the self-energy and vertex equations in diagrams. Using the earlier graphical notation for the bare and exact vertices from \S\ref{WyldSubsection1}, the self-energy, Eq. \eqref{BigSig}, can be presented as
\begin{equation}\label{OriginalMSR_SelfE}
\includegraphics[scale=1.0]{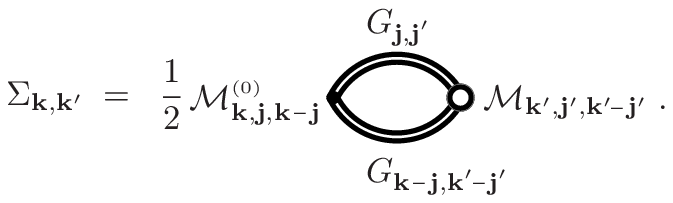}
\end{equation}
Similarly, the vertex diagram, Eq. \eqref{BigGam}, is
\begin{equation}
\includegraphics[scale=1.0]{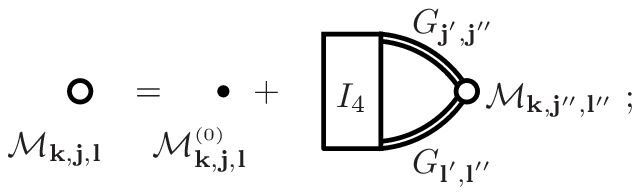}
\end{equation}
the quantity $I_4$ is used here in place of $\delta\bigma{k,j}{t,t'}/\delta\bigG{k',j'}{s,s'}$. This equation can be substituted into the vertex term of Eq. \eqref{OriginalMSR_SelfE}, giving 
\begin{equation}\label{BigMSR_SelfE}
\includegraphics[scale=1.0]{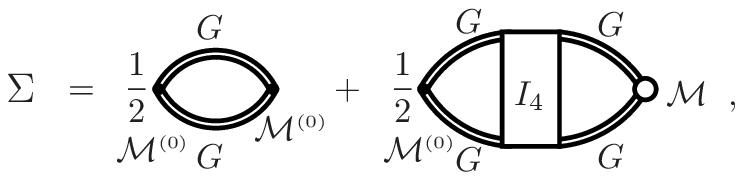}
\end{equation}
where wave-vector labels have been suppressed; 
the corresponding equation to this diagram is
\beq
\bigma{k,k'}{t,t'} = \frac{1}{2}\Totu{k,j,l}{}\bigG{j,j'}{s,s'}\bigG{l,l'}{s,s'}\Totu{k',j',l'}{}+\frac{1}{2}\Totu{k,j,l}{}\bigG{j,j'}{s,s'}\bigG{l,l'}{s,s'}\left[\frac{\delta\bigma{j',l'}{t,t'}}{\delta\bigG{m,n}{s,s'}}\right]\bigG{m,m'}{s,s'}\bigG{n,n'}{s,s'}\Gramma{k',m',n'}{3}.
\eeq
We introduce a diagram for the four-point term $I_4$ for convenience; it can be written as
\begin{equation}\label{IFour}
\includegraphics[scale=1.0]{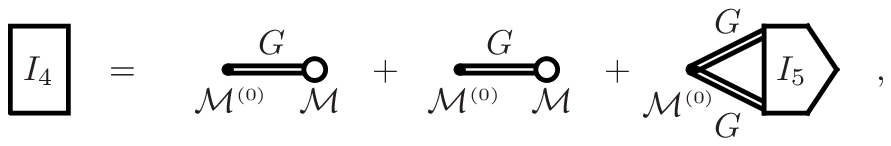}
\end{equation}
with the corresponding equation
\bey
I_4 & = & \frac{\delta\bigma{j',l'}{t,t'}}{\delta\bigG{m,n}{s,s'}}\nonumber\\
& = & \frac{\da}{\da\bigG{m,n}{s,s'}}\left(\frac{1}{2}\Totu{j',g,h}{}\bigG{g,g'}{s,s'}\bigG{h,h'}{s,s'}\Gramma{l',g',h'}{}\right)\nonumber\\
& = & \frac{1}{2}\Totu{j',m,h}{}\bigG{h,h'}{s,s'}\Gramma{l',n,h'}{} + \frac{1}{2}\Totu{j',g,m}{}\bigG{g,g'}{s,s'}\Gramma{l',g',n}{}\nonumber\\
& & + \;\; \frac{1}{2}\Totu{j',g,h}{}\bigG{g,g'}{s,s'}\bigG{h,h'}{s,s'}\frac{\da}{\da\bigG{m,n}{s,s'}}\left(\Gramma{l',g',h'}{}\right).
\eey
The term $I_5\equiv\da\Gramma{l',g',h'}{}/\da\bigG{m,n}{s,s'}$ has also been 
temporarily introduced. As the next steps are intermediate, we will briefly 
suppress labels. Inserting Eq. \eqref{IFour} into Eq. \eqref{BigMSR_SelfE} results 
in an expansion, 
\begin{equation}
\includegraphics[scale=1.0]{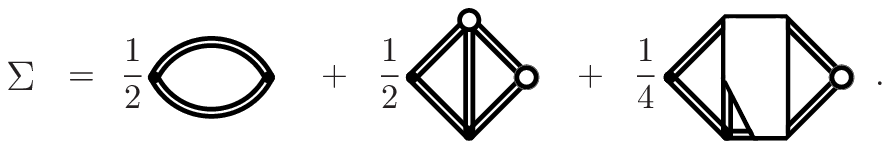}
\end{equation}
This can be further written with the bare vertices as
\begin{equation}\label{seeq}
\includegraphics[scale=1.0]{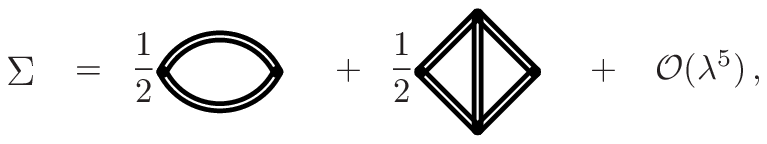}
\end{equation}
where we make the association of $\lambda$ with the bare vertex term $\Totu{}{}$ as was done in the previous section. 

Note that $\Totu{}{}$ is a $2\times2\times2$ tensor, and it can 
be shown \cite{Rose74,Krommes02} to be symmetric with three nonzero 
entries,
\beq\label{mbare}
\left[\Totu{k,j,l}{}\right]_{\scriptscriptstyle{++-}}=\left[\Totu{k,j,l}{}\right]_{\scriptscriptstyle{+-+}}=\left[\Totu{k,j,l}{}\right]_{\scriptscriptstyle{-++}}.
\eeq
Using this and noting that only $\left[\bigG{a,b}{}\right]_{\scriptscriptstyle{--}}$ vanishes, the self-energy is constrained to have only one zero component, $\left[\bigma{a,b}{}\right]_{\scriptscriptstyle{++}}$. The remaining terms on the $RHS$ of the self-energy equation Eq. \eqref{seeq} are found to be 
\begin{equation}\label{vertcor1}
\includegraphics[scale=1.0]{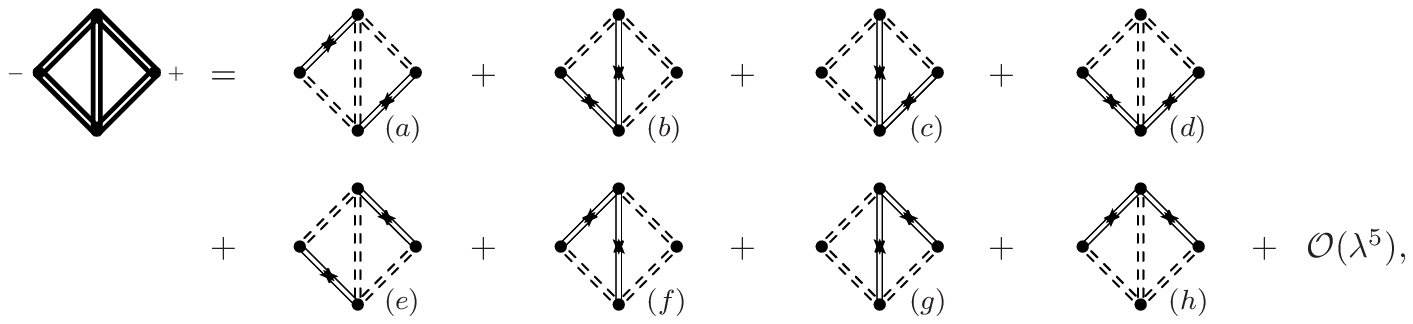}
\end{equation}
for the off-diagonal terms, and 
\begin{equation}\label{vertcor2}
\includegraphics[scale=1.0]{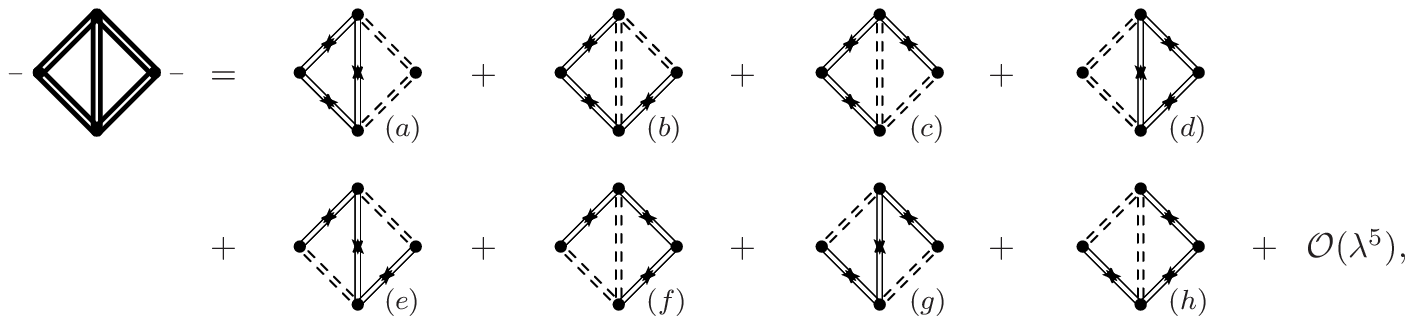}
\end{equation}
are the non-vanishing diagonal components. Both of these equations 
have been shown only to fourth-order in $\Totu{}{}$ as this is the extent 
of this study.

We have obtained the above results, Eqs. \eqref{vertcor1} and \eqref{vertcor2} 
from the
basic equations in MSR, and as we will see later in this Section,
they give a perturbation expansion that agrees with Wyld up to fourth order.
However they never actually calculated in their paper the fourth order 
self-energy contributions as we have in 
Eqs. \eqref{vertcor1} and \eqref{vertcor2}.
Instead MSR recast their formalism into a simplified
diagramatic approach and from this their perturbation expansion was
developed.

The MSR diagramatic approach began by writing
the fourth order
contributions to the self-energy in terms of
two Green's function lines convoluted with vertex functions.
They wrote the non-vanishing elements of the self-energy tensor
through the diagrammatic equations
\begin{equation}\label{NewMSR_SelfE_1}
\includegraphics[scale=1.0]{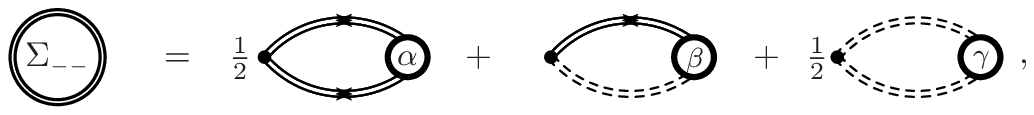}
\end{equation}
\begin{equation}\label{NewMSR_SelfE_3}
\includegraphics[scale=1.0]{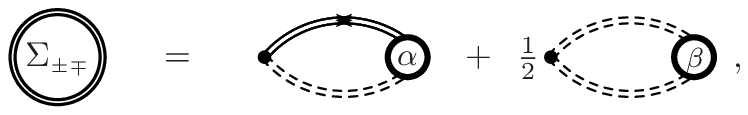}
\end{equation}
where $\alpha$, $\beta$, and $\gamma$ label the three vertex functions, 
which MSR claimed to be
\begin{equation}\label{NewMSR_Vertices1}
\includegraphics[scale=1.0]{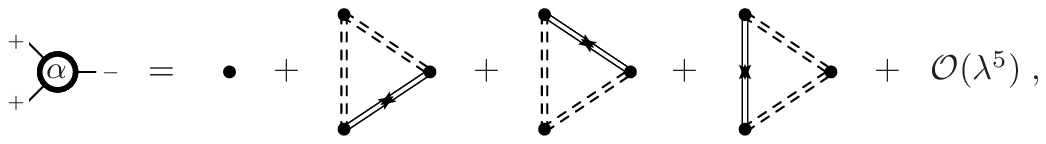}
\end{equation}
\begin{equation}
\label{MSR_Vertices2}
\includegraphics[scale=1.0]{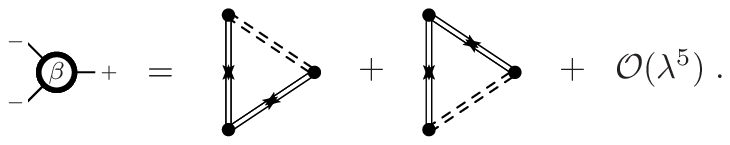}
\end{equation}
\begin{equation}\label{NewMSR_Vertices3}
\includegraphics[scale=1.0]{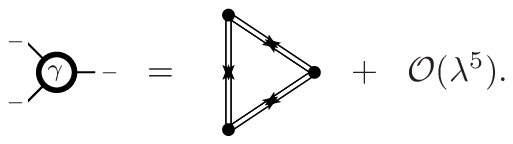}
\end{equation}
These three terms correspond to the non-vanishing elements of $\mathcal{M}$, 
which
are $[\mathcal{M}]_{\scriptscriptstyle{++-}}$, $[\mathcal{M}]_{\scriptscriptstyle{--+}}$,
and $[\mathcal{M}]_{\scriptscriptstyle{---}}$, and
their permutations. 
We have displayed these vertex functions noting specifically that for 
a given combination of indices, the $\alpha$-vertex function has 
four diagrams, the $\beta$-vertex has two diagrams, and the $\gamma$-vertex 
only one. 
The expressions on the $LHS$ in the above equations indicate how vertex diagrams attach to the self-energy terms. Permuting indices leaves the vertices unchanged, for example 
$[\mathcal{M}]_{\scriptscriptstyle{--+}}=[\mathcal{M}]_{\scriptscriptstyle{-+-}}=[\mathcal{M}]_{\scriptscriptstyle{+--}}$, and diagrammatically this corresponds to in-plane rotations and  reflections about a line connecting a vertex and the midpoint of an opposing edge. Such permutations are trivial for the $\alpha$- and $\gamma$-vertex functions where any  operation returns the same diagrams given in the above equations. The $\beta$-vertices require these permutations to generate the necessary self-energy terms. For example, $\Sigma_{\scriptscriptstyle{\mp\pm}}$ uses Eq. \eqref{MSR_Vertices2} as it is shown; however, $\Sigma_{\scriptscriptstyle{--}}$  requires a $120^{\scriptstyle{\circ}}$ counterclockwise rotation, 
$[\mathcal{M}]_{\scriptscriptstyle{--+}}\to[\mathcal{M}]_{\scriptscriptstyle{+--}}$, for diagrams Eq. \eqref{vertcor2}$(f)$ and $(e)$, respectively, whereas a $120^{\scriptstyle{\circ}}$ clockwise rotation, $[\mathcal{M}]_{\scriptscriptstyle{--+}}\to[\mathcal{M}]_{\scriptscriptstyle{-+-}}$, is needed for Eq. \eqref{vertcor2}$(g)$ and $(h)$, respectively. In this manner, one can confirm that all the diagrams in Eqs. \eqref{vertcor1} and \eqref{vertcor2} are produced.
We will examine now in close detail how the formalisms of MSR
and Wyld lead to
complete agreement at fourth order.
Inserting the vertex corrections, 
Eqs. \eqref{NewMSR_Vertices1}, \eqref{MSR_Vertices2},
and \eqref{NewMSR_Vertices3}
into their 
respective positions in the self-energy diagrams, Eq. \eqref{NewMSR_SelfE_1} 
and Eq. \eqref{NewMSR_SelfE_3}, gives the self-energy diagrams to fourth-order. 
Then, the expanded self-energies are inserted into Eq. \eqref{MSR_1R} 
to obtain from the MSR formalism the propagator diagrams to fourth-order,
\begin{equation}\label{MSR_4_1}
\includegraphics[scale=1.0]{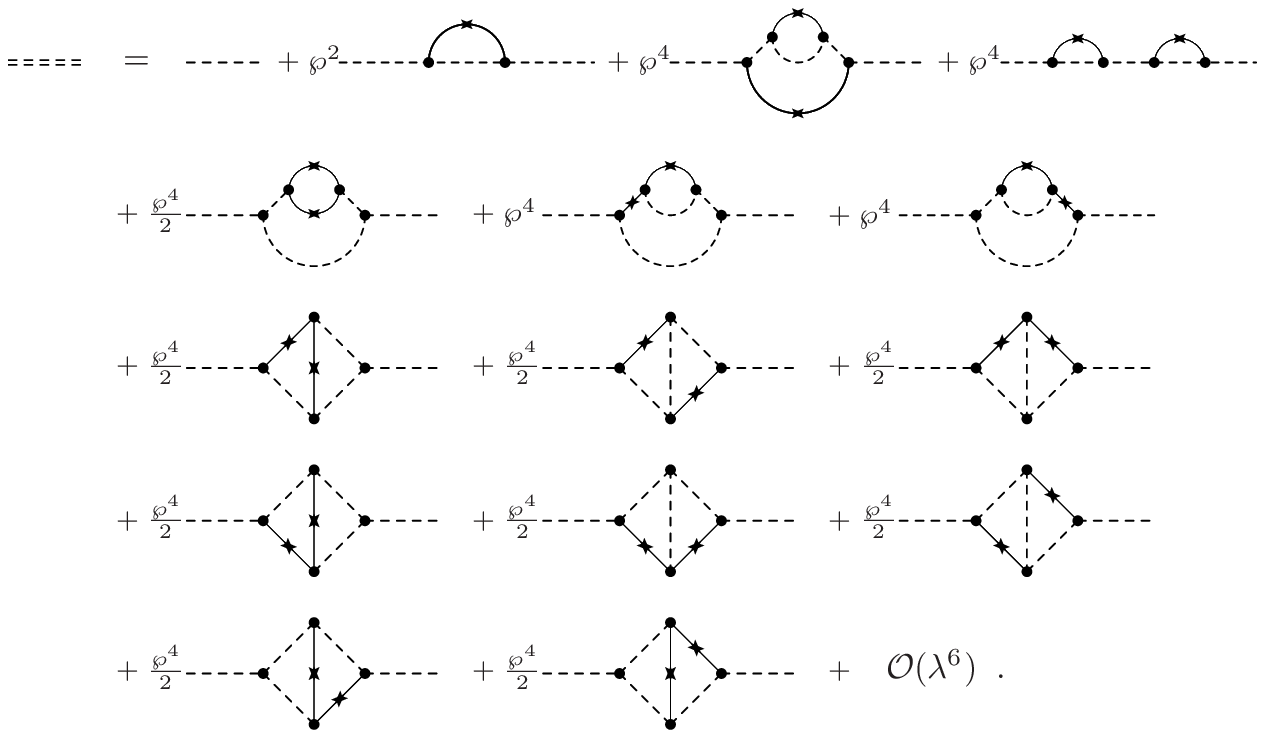}
\end{equation}
A factor $\wp$ has been introduced for each vertex since 
this overall factor differs
between MSR and Wyld.
Using this propagator expansion, the terms in Eq. \eqref{MSR_1C_2} for the 
exact correlator may now be determined. Those terms obtained from 
the forcing function are
\begin{equation}\label{MSR_4_2}
\includegraphics[scale=1.0]{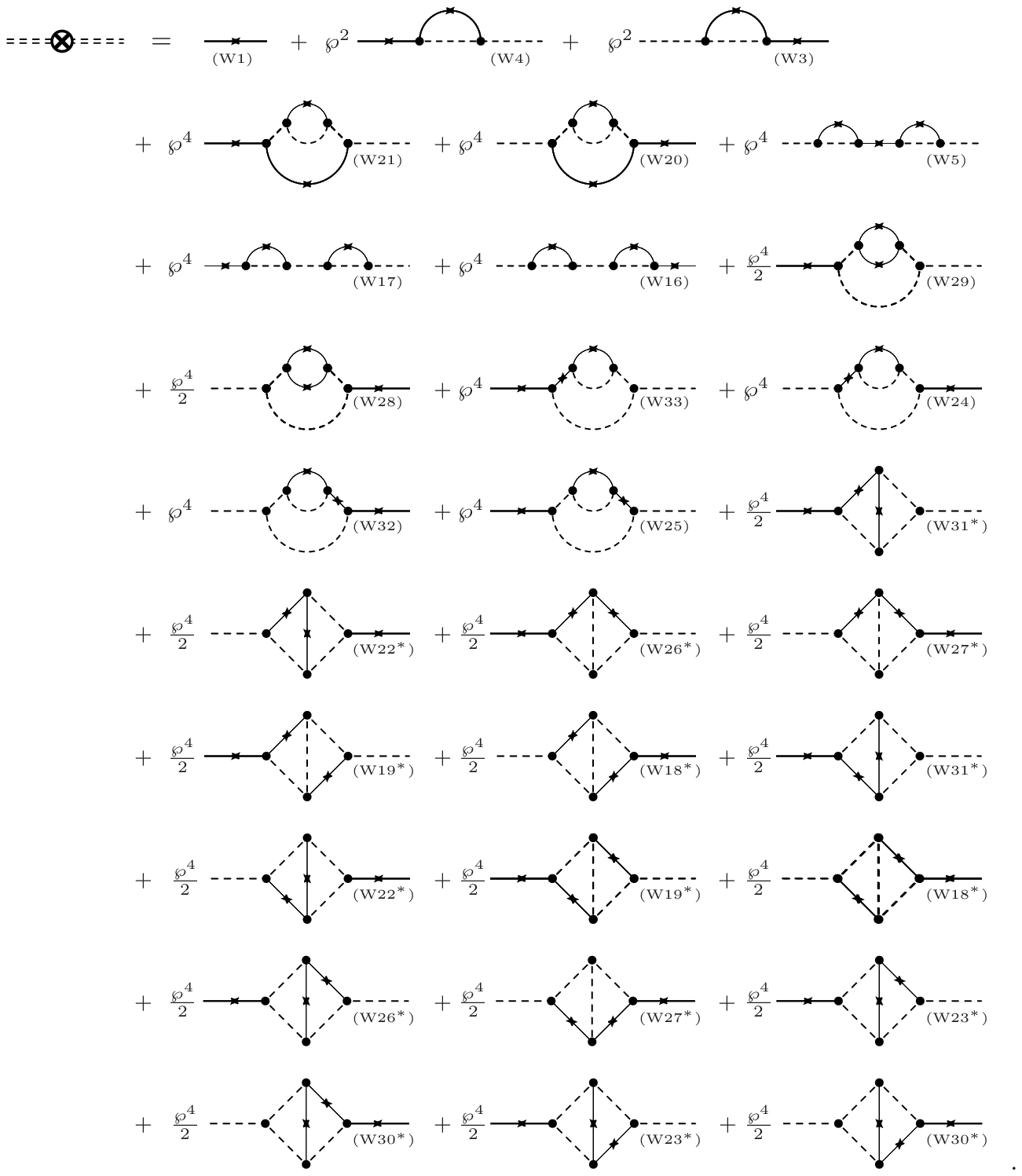}
\end{equation}
The labels (W$n$) correspond to Wyld's diagrams as seen in Fig.~\ref{Correlation_1_3}. Labels containing an asterisk denote diagrams with half the weighting of their Wyld counterparts; however there are always two such diagrams and their sum gives the correct weighting. These diagrams are symmetric when reflected about a horizontal line and in Wyld's formalism such diagrams are equivalent. 

Those diagrams representing the self-energy interaction are expressed by 
\begin{equation}\label{MSR_4}
\includegraphics[scale=1.0]{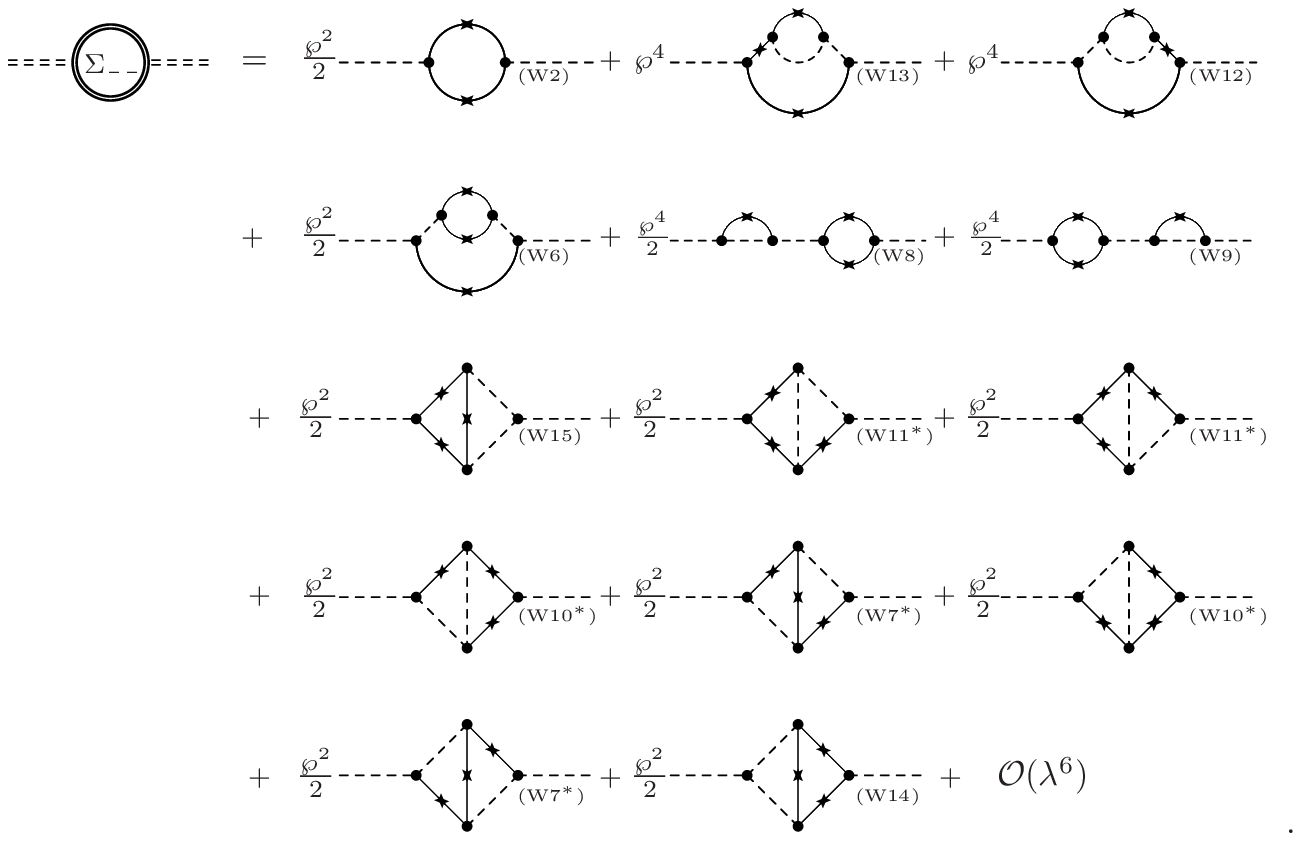}
\end{equation}
Combining Eq. \eqref{MSR_4_2} and Eq. \eqref{MSR_4} together gives the diagrammatic 
equation for the exact correlator function expanded to fourth-order in 
the bare vertex as obtained by the MSR formalism. There are 44 diagrams 
counted in these two equations, confirming a statement made in the MSR paper.
Some of these 44 diagrams are repetitions of the same diagram.
If these diagrams are then combined taking appropriate account
for weight factors, they yield the 33 diagrams in the Wyld
perturbation expansion in Fig. \ref{Correlation_1_3}. 
Comparing these two equations 
to the diagrams given in Fig.~\ref{Correlation_1_3}, it can be concluded 
that the MSR formalism gives the primitive correlator expansion of Wyld 
provided the factor $\wp=2$. 
\section{Comparison}\label{Comparison}

The renormalized diagrammatic expressions for homogeneous isotropic 
turbulence have been presented here according to the formalisms of 
Wyld (sections \S\S\ref{WyldSubsection1}--\ref{WyldSubsection2}) and 
Martin, Siggia, and Rose (\S \ref{MSRSubsection1}).  Comparisons 
are made between
the Improved Wyld-Lee formalism presented in 
section \S\ref{WyldSubsection2} to expressions for the 
self-energy and vertex diagrams derived by MSR in the previous
section.   Ultimately we show that both formalisms are equivalent
to fourth order.

To compare the diagram equations as obtained in their respective formalisms, some rearrangement of terms is necessary. Looking again at the propagator expression of Wyld in Eq. \eqref{LeeKW_prop}, this equation can be written in such a way as to anticipate the MSR form of the propagator with the self-energy term
\begin{equation}\label{WyldMSR_Comp2}
\includegraphics[scale=1.0]{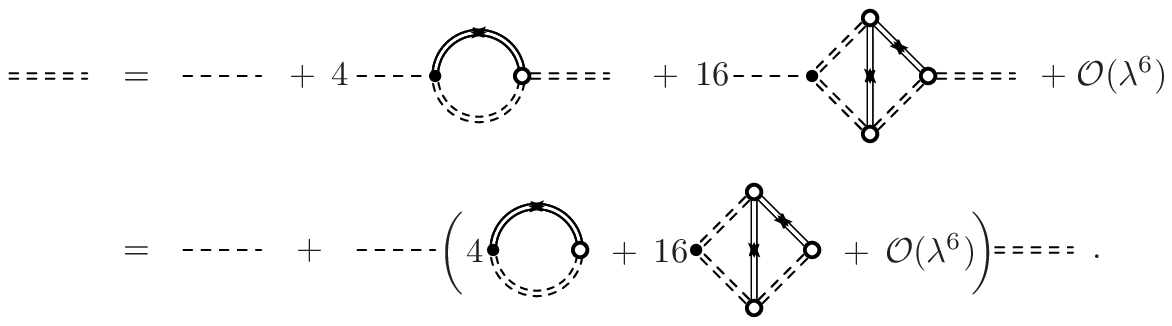}
\end{equation}
Noting the reflection-symmetry about
the horizontal line of Wyld diagrams, the fourth-order term 
in Eq. \eqref{WyldMSR_Comp2} can be written as a sum of two terms,
\begin{equation}\label{WyldMSR_Comp}
\begin{picture}(320,42)
\put(8,0){\includegraphics[scale=0.9]{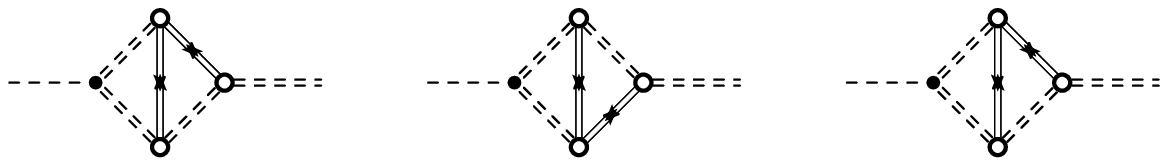}}
\put( 2,17.){$2$} \put(100,17.){$=$} \put(209,17.){$+$}
\end{picture}.
\end{equation}
This allows the above equation for the propagator, Eq. \eqref{WyldMSR_Comp2}, to be written using two of the MSR-vertex terms, $\mathcal{M}^{(\al)}$ and $\mathcal{M}^{(\ba)}$ (see Eq. \eqref{NewMSR_Vertices1} and Eq. \eqref{MSR_Vertices2}): 
\begin{equation}\label{WyldMSR_Comp}
\includegraphics[scale=1.0]{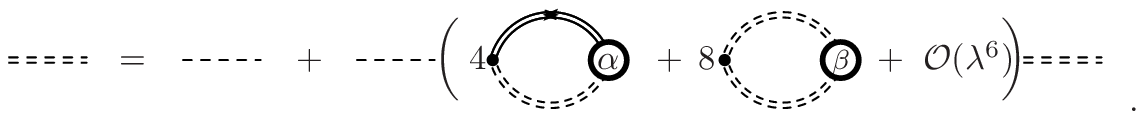}
\end{equation}
In all further diagrams of both formalisms, diagrams that are equivalent 
through this symmetry will be combined. The ``$\order{6}$'' has been kept 
explicit as the above only applies to fourth-order; the next order has 
not been analysed. Substituting the off-diagonal self-energy term,
Eq. \eqref{NewMSR_SelfE_3}, in place of the terms in the brackets of 
Eq. \eqref{WyldMSR_Comp}, we find that the MSR propagator given in 
Eq. \eqref{MSR_1R} is recovered,
\begin{equation}\nonumber
\begin{picture}(196,30)
\put(0,0){\includegraphics[scale=1.0]{./MSR_R.eps}}
\put(146,11){$\Sigma_{\scriptscriptstyle{\pm\mp}}$}
\put(40,12){=\qquad\qquad\quad\,\! +}
\end{picture}.
\end{equation}
Before continuing with Wyld, it will be helpful to rewrite the diagram equation for the exact MSR-correlator, seen in Eq. \eqref{MSR_1C_2} and reproduced here,
\begin{equation}\nonumber
\begin{picture}(226,32)
\put(0,0){\includegraphics[scale=1.0]{./MSR_CC.eps}}
\put(173,11.5){$\Sigma_{\scriptscriptstyle{--}}$}
\put(44,13){=\qquad\qquad\quad\qquad\; +}
\end{picture}.
\end{equation}
Substituting the propagator equation given above (from Eq. \eqref{MSR_1R}) for the exact propagator on the left of the force-force correlation, $\crln{\mathbf{f\otimes f}}$ and the self-energy term, $\Sigma_{\scriptscriptstyle{--}}$, the correlator equation becomes 
\begin{equation}\label{MSR_DIA_1}
\includegraphics[scale=1.0]{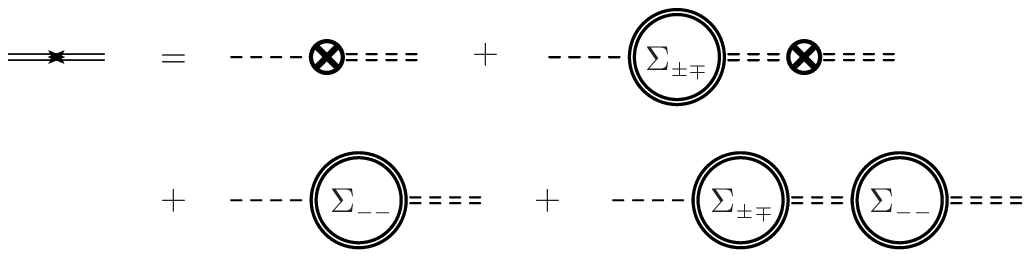}
\end{equation}
However, in making the comparison with Wyld, it will be better to use Eq. \eqref{MSR_DIA_1}. To fourth-order, the self-energy terms can be written explicitly as 
\begin{equation}\label{MSR_SelfEVerts_ba}
\includegraphics[scale=1.0]{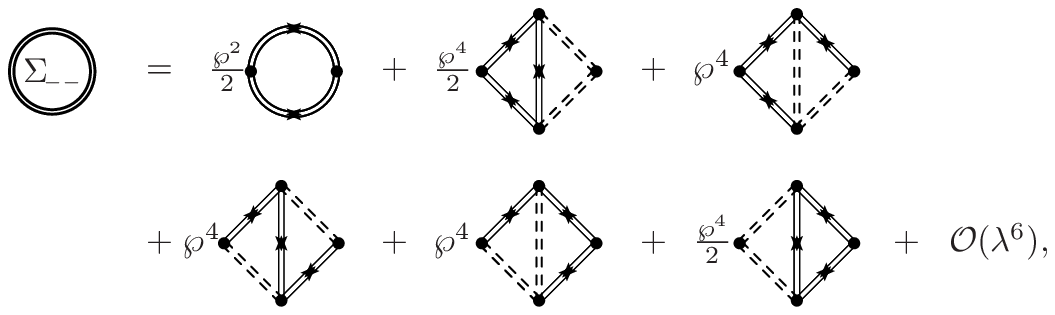}
\end{equation}
\begin{equation}\label{MSR_SelfEVerts_b}
\includegraphics[scale=1.0]{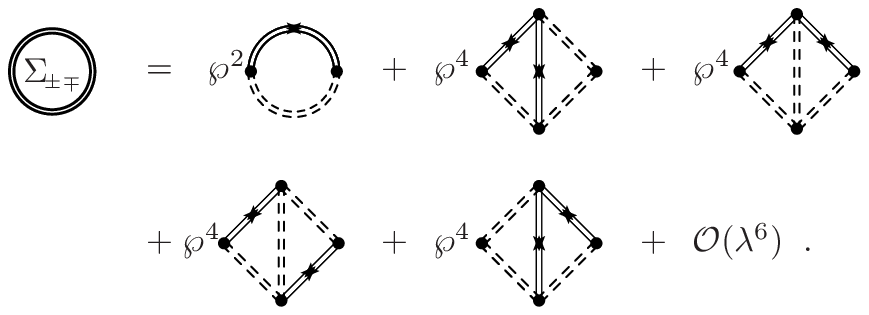}
\end{equation}
The factor $\wp$, associated with each vertex, has been included for 
convenience.  Note that independent of MSR, these terms had been 
derived at about the same time as them for the case of wave 
turbulence in \cite{Zakharov75}.
The terms with only one self-energy component are given as 
\begin{equation}\label{KWMSR_SelfEVerts_a}
\includegraphics[scale=1.0]{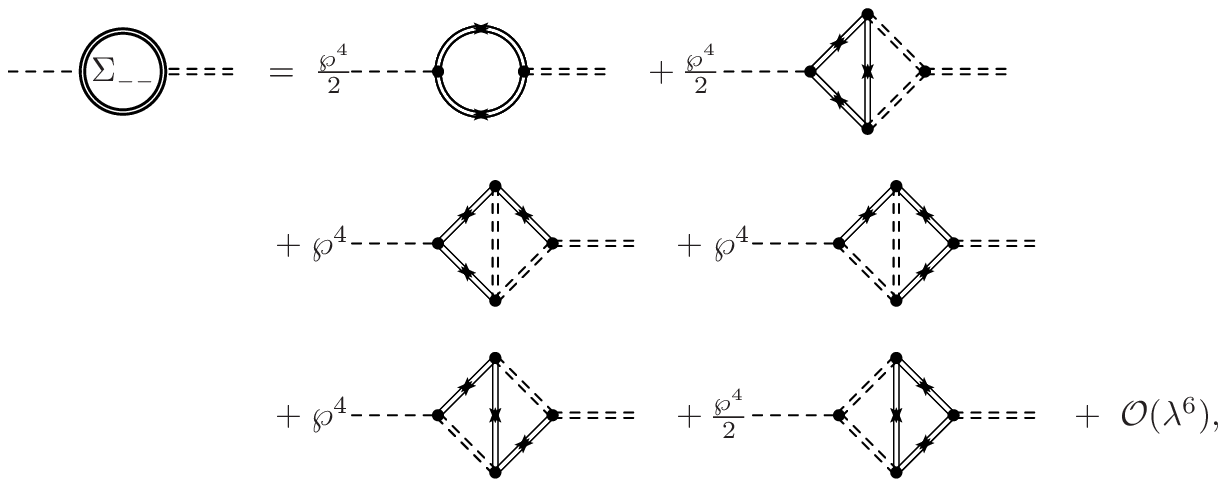}
\end{equation}
and
\begin{equation}\label{KWMSR_SelfEVerts_b}
\includegraphics[scale=1.0]{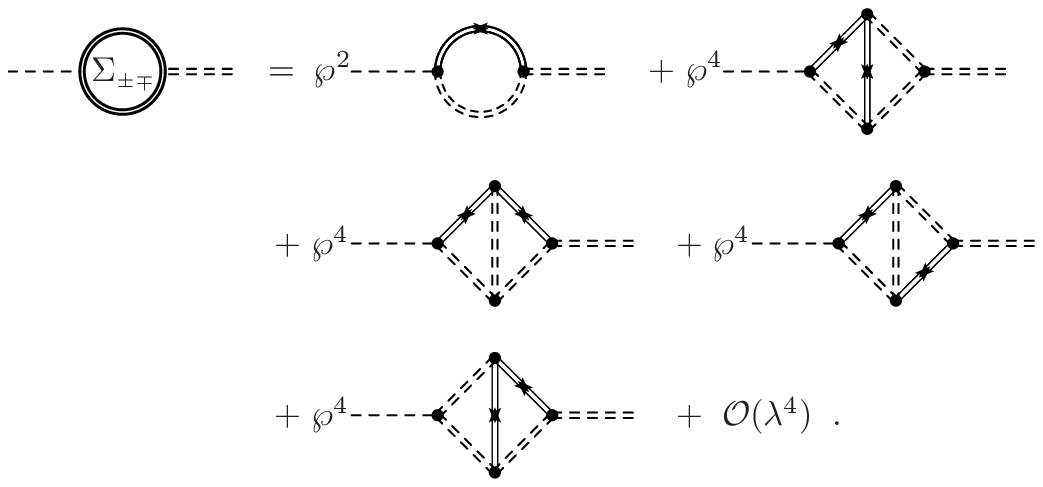}
\end{equation}
The term with both $\Sigma_{\scriptscriptstyle{--}}$ and $\Sigma_{\scriptscriptstyle{\pm\mp}}$ will not be expanded. This term can be combined with Eq. \eqref{KWMSR_SelfEVerts_b} to give
\begin{equation}\label{MSR_Corr_0th}
\begin{picture}(360,32)
\put(0,0){\includegraphics[scale=1.0]{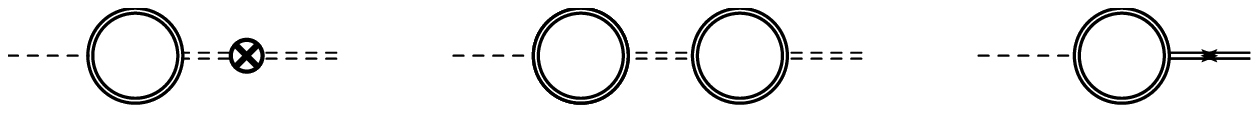}}
\put(29,12){$\Sigma_{\!\scriptscriptstyle{\pm\mp}}$}
\put(108,13){$+$}
\put(158,12){$\Sigma_{\!\scriptscriptstyle{\pm\mp}}$}
\put(203,12){$\Sigma_{\!\scriptscriptstyle{--}}$}
\put(260,13){$=$}
\put(314,12){$\Sigma_{\!\scriptscriptstyle{\pm\mp}}$}
\end{picture}.
\end{equation}
Using the expressions Eq. \eqref{KWMSR_SelfEVerts_a}-\eqref{MSR_Corr_0th}, these equations can be substituted back into Eq. \eqref{MSR_DIA_1}, 
\begin{equation}\label{KWMSR_corr_ALL}
\includegraphics[scale=1.0]{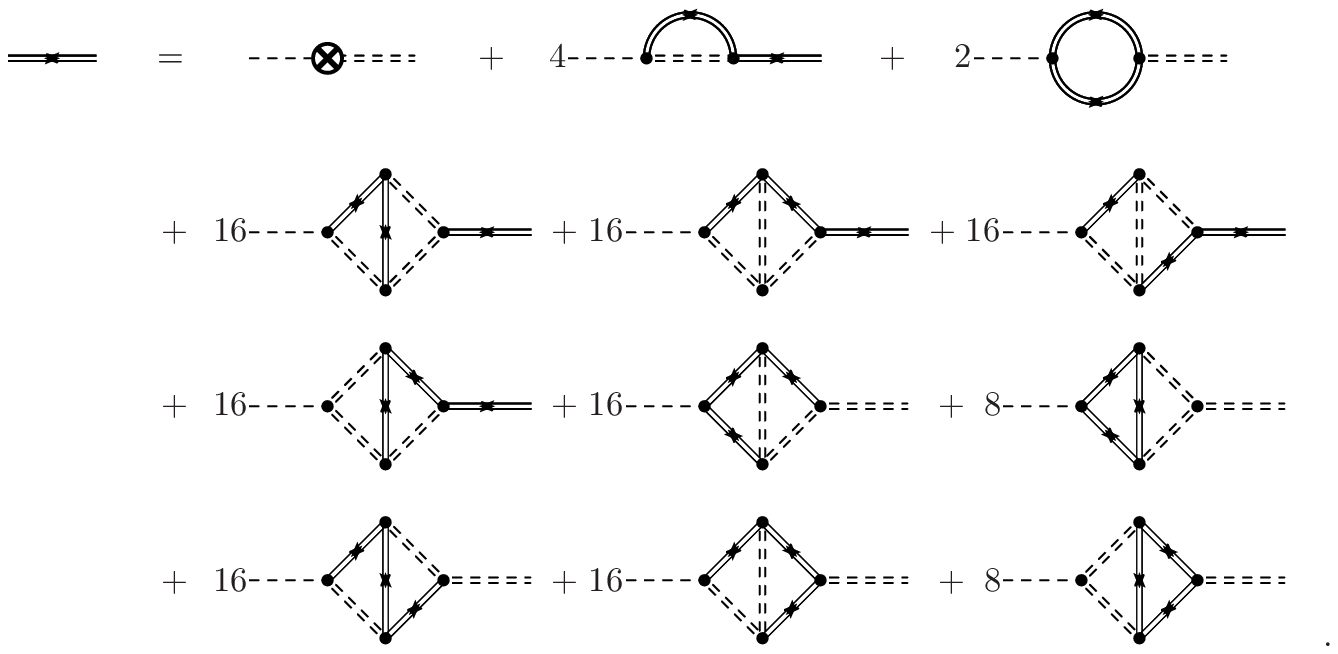}
\end{equation}
The term $\wp=2$ has been used, which will allow the 
comparison with Wyld. To make a comparison with this correlator, 
some adjustment is required for the Wyld correlator. 
The equation for the correlator given in Eq. \eqref{LeeKW_corr_full},
\begin{equation}\label{LeeKW_corr_fullo}
\includegraphics[scale=1.0]{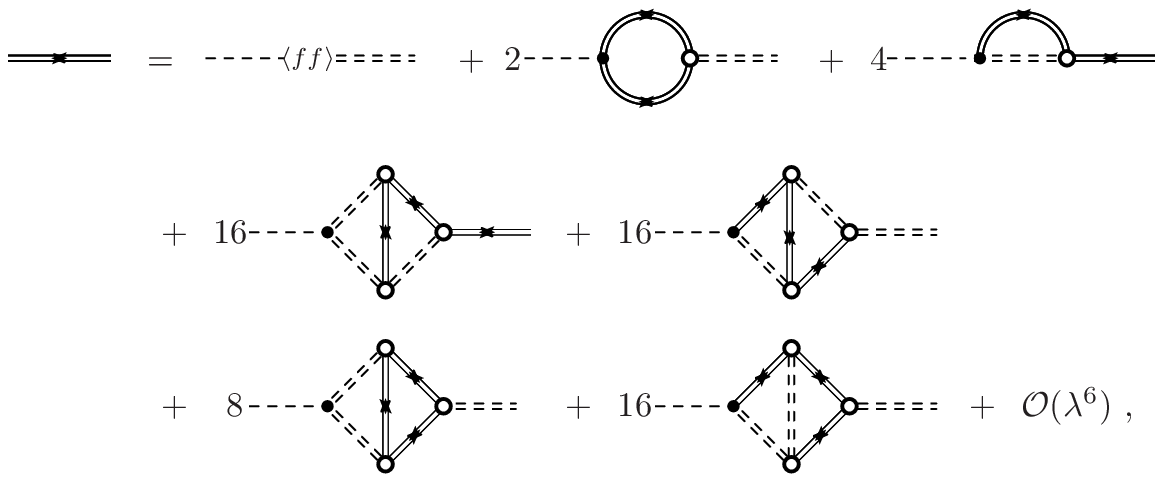}
\end{equation}
can be expanded using the renormalized vertex expansion of Eq. \eqref{exact_vert_func} and truncated to fourth-order, leaving
\begin{equation}\label{KW_corr_ALL}
\includegraphics[scale=1.0]{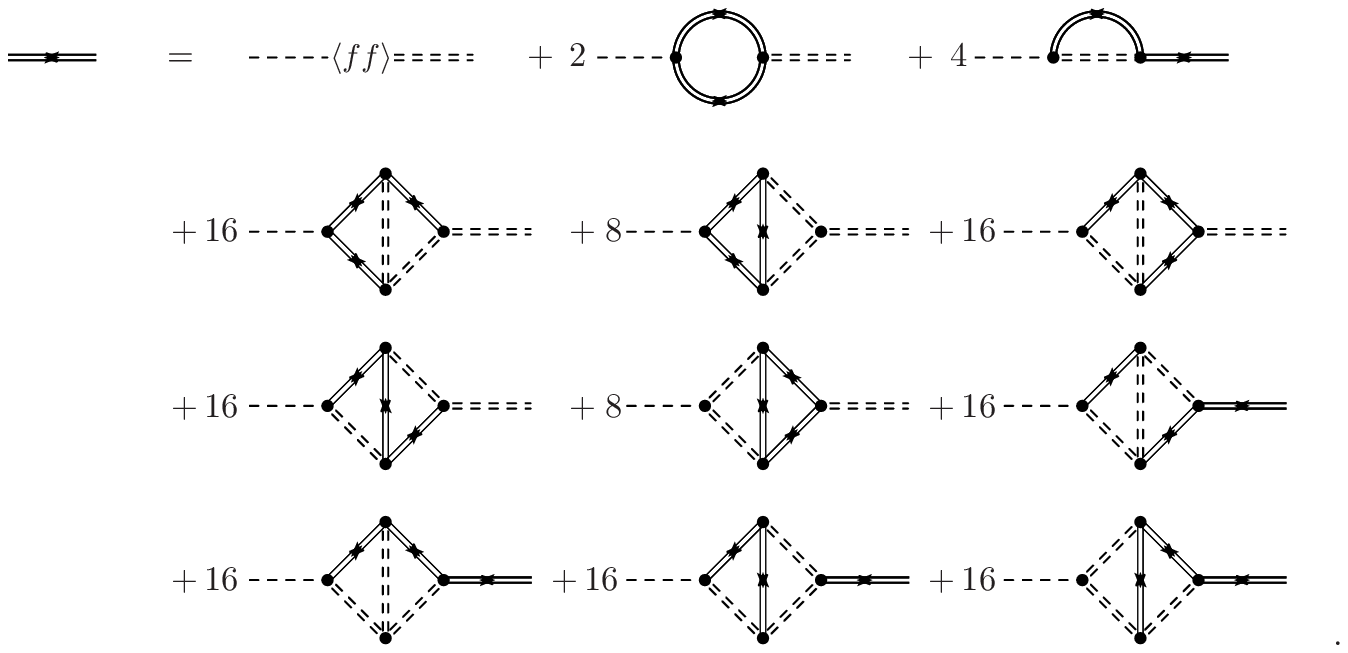}
\end{equation}
This equation can be readily compared to Eq. \eqref{KWMSR_corr_ALL}. The double-force correlation term in Wyld can be connected with that given in MSR (see Eq. \eqref{MSR_1X}),
\begin{equation}
\begin{picture}(50,11)
\put(35,-2){\includegraphics[scale=0.9]{./MSR_FF.eps}}
\put(0,2.){$\scriptstyle{\crln{ff}}$} \put(19,1.){$\,=$}
\end{picture}
\end{equation}.

The equations Eq. \eqref{KWMSR_corr_ALL} and Eq. \eqref{KW_corr_ALL} are equivalent. 
It can be said that the results produced here by both formalisms are equivalent 
to fourth-order. This raises questions on the claims by MSR and Kraichnan over 
the incorrect treatment of the vertex renormalization by Wyld. Their claim was 
that Wyld was missing certain classes of vertex corrections, 
namely Eq. \eqref{NewMSR_Vertices3} and Eq. \eqref{MSR_Vertices2}. 
This is true, but Wyld's resummation procedure is also different
from MSR.  In the Wyld approach, at every order in the perturbation
expansion, he adds what he calls irreducible diagrams.
Notice in the original Wyld formulation
his correlation equation \eqref{w-3_ii} had two types of irreducible
diagrams; one can compare to the corresponding MSR equation
\eqref{NewMSR_SelfE_1} and see the difference. 
In making this comparison, importantly
note that
the leftmost vertex in the Wyld case is
also renormalized, which again differs from MSR.
Nevertheless, as we saw on comparing Wyld and MSR, that up
to fourth order, they lead to
equivalent perturbation
expansions. We made a modification to the original
Wyld procedure, in order to have a single procedure
for treating both the propagator and correlator.
For this, we modified the correlator equation of Wyld to
Eq. \eqref{LeeKW_corr_full}, in which, just like for 
the propagator, the leftmost
vertex and propagator remain unrenormalized.  However the basic
principle adopted by Wyld to include certain irreducible diagrams
at each order in the perturbation expansion is still
adhered to in our Improved Wyld-Lee perturbation theory.
Within this approach, again we explicitly confirmed
that Eq. \eqref{LeeKW_corr_full} 
agreed with MSR to fourth order.


\section{Conclusion}\label{Conclusion}
The formalisms of Wyld and Martin, Siggia, and Rose have been presented in detail is this paper. 
Specific attention has been given to the diagrammatic interpretations 
of both formalisms, as this is where confusion can arise when
making comparison.

A new derivation of the propagator and correlator
for the Wyld formalism has been 
given, that obtains the corresponding expressions proposed by Lee. 
The main feature of this resummation procedure is that the leftmost propagator 
and leftmost vertex remain unrenormalized.  This procedure, when
added to the original Wyld formalism, we named here
as the Improved Wyld-Lee Renormalized Perturbation Theory,
which in particular comprises Eqs. \eqref{LeeKW_corr_full},
\eqref{LeeKW_prop}, and \eqref{W_vertex_expansion}.
Applying this method for the propagator, one directly obtains the DIA 
results when truncated to lowest nontrivial order,
just as one does with the MSR formalism.

Furthermore, equations of the Improved Wyld-Lee Formalism
for the exact correlator \eqref{LeeKW_corr_full} 
and propagator \eqref{LeeKW_prop}
have been shown to agree 
with the corresponding MSR 
equations \eqref{NewMSR_SelfE_1} and \eqref{NewMSR_SelfE_3} respectively.
What we have pointed out here is that
Wyld had a different procedure for expressing his basic equations,
involving what he called irreducible terms. Accounting
for these, we illustrated how the Wyld procedure agrees with
MSR up to fourth order.
We had included the factor $\wp$ in the MSR diagrams, 
\eqref{KWMSR_SelfEVerts_a} and Eq. \eqref{KWMSR_SelfEVerts_b}, to account 
for a weighting factor. We found that setting $\wp=2$ gives the proper 
Wyld weightings but this cannot be known {\it a priori} by the 
MSR procedure. 
However our corrections do not fundamentally change Wyld's approach
of classifying certain irreducible diagrams and accounting for
them in his perturbation expansion. It is this procedure that makes
his approach different from MSR, and due to that,
just a simple comparison of the vertex functions in both formalisms
is insufficient in making a comparison of
the two formalisms.
It is therefore the conclusion of this work that both 
formalisms are equivalent. 
This means that the Wyld formalism, with the diagrammatic 
resummation used here, produces equations for the exact correlator
and propagator 
that are the same as those obtained using the 
formalism of MSR.

\section*{Acknowledgements}

AB was funded by STFC, MS was supported by the 
Deutsche Forschungsgemeinschaft (͑DFG), and WDM acknowledges the support 
of an Emeritus Fellowship from the Leverhulme Trust.



\begin{thebibliography}{10}

\bibitem{Davidson04}
P.~A.~Davidson.
\newblock {\em Turbulence: an introduction for scientists and engineers}.
\newblock (Oxford University Press, Oxford, 2004)

\bibitem{Wyld61}
H.~W. Wyld.
\newblock {\em Ann. Phys.} {\bf 14}, 143 (1961).

\bibitem{MaSiRo73}
P.~C. Martin{,} E. D. Siggia{,} H.~A. Rose.
\newblock {\em Phys. Rev. A} {\bf 8}, 423 (1973).

\bibitem{Millionshchikov41}
M.~D. Millionshchikov.
\newblock {\em Dokl. Akad. Nauk SSSR} {\bf 32} (1941).

\bibitem{PrRe54}
I.~Proudman and W.~H. Reid.
\newblock {\em Phil. Trans. Roy. Soc. London Ser. A} {\bf 247}, 163 (1954).

\bibitem{Ogura63}
Y.~Ogura.
\newblock {\em J. Fluid Mech.} {\bf 16}, 33 (1963).

\bibitem{Lesieur90}
M.~Lesieur.
\newblock {\em {Turbulence in fluids}}.
\newblock (Kluwer Academic Publishers, Dordrecht, 1990).

\bibitem{Leslie73}
D.~C. Leslie.
\newblock {\em Developments in the Theory of Turbulence}.
\newblock (Clarendon Press, Oxford, 1973).

\bibitem{McComb90}
W.~D. McComb.
\newblock {\em The {P}hysics of {F}luid {T}urbulence}.
\newblock (Clarendon Press, Oxford, 1990).

\bibitem{Kraichnan57}
R.~H. Kraichnan.
\newblock {\em Phys. Rev.} {\bf 107}, 1485 (1957).

\bibitem{Kraichnan58}
R.~H. Kraichnan.
\newblock {\em Phys. Rev.} {\bf 109}, 1407 (1958).

\bibitem{Kraichnan59}
R.~H. Kraichnan.
\newblock {\em J. Fluid Mech.} {\bf 5}, 497 (1959).

\bibitem{Beran68}
M.~Beran.
\newblock {\em Statistical Continuum Theories}.
\newblock (Interscience Publishers, New York, NY, 1968).

\bibitem{McComb95}
W.~D. McComb.
\newblock {\em {R}ep. {P}rog. {P}hys.} {\bf {\bf 58}}, 1117 (1995).

\bibitem{KiGo97}
S.~Kida and S.~Goto.
\newblock {\em J. Fluid Mech.} {\bf 345}, 307 (1997).

\bibitem{Krommes02}
J.~A. Krommes.
\newblock {\em Phys. Rep.} {\bf 360}, 1 (2002).
\newblock Section 6.

\bibitem{Kraichnan64}
R.~H. Kraichnan.
\newblock {\em Phys. Fluids} {\bf 7}, 1030 (1964).

\bibitem{Kraichnan66}
R.~H. Kraichnan.
\newblock {\em Phys. Fluids} {\bf 9}, 1728 (1966).

\bibitem{Nakano72}
T.~Nakano.
\newblock {\em Ann. Phys.} {\bf 73}, 326 (1972).

\bibitem{McComb74}
W.~D. McComb.
\newblock {\em{Proc. Roy. Soc. Edinburgh A}} {\bf 72}, 18 (1974).

\bibitem{McComb74-2}
W.~D. McComb.
\newblock {\em{J. Phys. A}} {\bf 7}, 632 (1974).

\bibitem{McComb76}
W.~D. McComb.
\newblock {\em{J. Phys. A}} {\bf 9}, 179 (1976).

\bibitem{McComb78}
W.~D. McComb.
\newblock {\em{J. Phys. A}} {\bf 11}, 613 (1978).

\bibitem{McComb84}
W.~D. McComb and V. Shanmugasundaram.
\newblock {\em{J. Fluid Mech.}} {\bf 143}, 95 (1984) .

\bibitem{McComb89}
W.~D. McComb, V. Shanmugasundaram and P. Hutchinson.
\newblock {\em{J. Fluid Mech.}} {\bf 208}, 91 (1989).

\bibitem{McFiSh92}
W.~D. McComb, M. J. Filipiak and V. Shanmugasundaram.
\newblock {\em{J. Fluid Mech.}} {\bf 245}, 279(1992). 

\bibitem{ObMcQu01}
M. Oberlack, W.~D. McComb and A.P. Quinn.
\newblock {\em{Phys. Rev. E}} {\bf 63}, 026308 (2001).

\bibitem{McQu03}
W.~D. McComb and A.~P. Quinn.
\newblock {\em{Physica A}} {\bf 317}, 487 (2003).

\bibitem{KiMc04}
K. Kiyani and W.~D. McComb.
\newblock {\em{Phys. Rev. E}} {\bf 70}, 066303 (2004).

\bibitem{McKi05}
W.~D. McComb and K. Kiyani.
\newblock {\em{Phys. Rev. E}} {\bf 72}, 016309 (2005).

\bibitem{Kraichnan65}
R.~H. Kraichnan.
\newblock {\em Phys. Fluids} {\bf 8}, 575 (1965).

\bibitem{Orszag06}
S.~A. Orszag.
\newblock {\em J. Fluid Mech.} {\bf 41}, 363 (2006).

\bibitem{Kaneda81}
Y.~Kaneda.
\newblock {\em J. Fluid Mech.} {\bf 107}, 131 (1981).

\bibitem{Kaneda86}
Y.~Kaneda.
\newblock {\em Phys. Fluids} {\bf 27}, 701 (1986).

\bibitem{FrDa00}
J.~S. Frederiksen and A.~G. Davies.
\newblock {\em Geophys. Astrophys. Fluid Dyn.} {\bf 92}, 197 (2000).

\bibitem{OKaFr04}
T.~J. O'Kane and J.~S. Frederiksen.
\newblock {\em J. Fluid Mech.} {\bf 504}, 133 (2004).

\bibitem{FrDan04}
J.~S. Frederiksen and A.~G. Davies.
\newblock {\em Geophys. Astrophys. Fluid Dyn.} {\bf 98}, 203 (2004).

\bibitem{FrOKa05}
J.~S. Frederiksen and T.~J. O'Kane.
\newblock {\em J. Fluid. Mech.} {\bf 539}, 137 (2005).

\bibitem{Hopf52}
E.~Hopf.
\newblock {\em J. Rat. Mech. Anal.} {\bf 1}, 87 (1952).

\bibitem{Edwards64}
S.~F.~Edwards.
\newblock {\em J. Fluid. Mech.} {\bf 18}, 239 (1964).

\bibitem{Herring65}
J.~R. Herring.
\newblock {\em Phys. Fluids} {\bf 8}, 2219 (1965).

\bibitem{Herring66}
J.~R. Herring.
\newblock {\em Phys. Fluids} {\bf 9}, 2106 (1966).

\bibitem{Teodorovich94}
E.~V. Teodorovich.
\newblock {\em Fluid Dyn.} {\bf 29}, 770 (1994).

\bibitem{Lee65}
L.~L. Lee.
\newblock {\em Ann. Phys.} {\bf 32}, 292 (1965).

\bibitem{PeSc95}
M.~E. Peskin and D.~V. Schroeder.
\newblock {\em {An Introduction to Quantum Field Theory}}.
\newblock (Addison-Wesley, Reading, MA, 1995).

\bibitem{Schwinger51a}
J.~Schwinger.
\newblock {\em Proc. Natl. Acad. Sci. USA} {\bf 37}, 452 (1951).

\bibitem{Schwinger51b}
J.~Schwinger.
\newblock {\em Proc. Natl. Acad. Sci. USA} {\bf 37)}, 455 (1951).

\bibitem{Schwinger51c}
J.~Schwinger.
\newblock {\em Phys. Rev.} {\bf 82}, 914 (1951).

\bibitem{Jensen81}
R.~V. Jensen.
\newblock {\em J. Stat. Phys.} {\bf 25}, 183 (1981).

\bibitem{Thacker97}
W.~D.~Thacker
\newblock {\em J. Math. Phys.} {\bf 38}, 300 (1997).

\bibitem{Rose74}
H.~A. Rose.
\newblock PhD Thesis, Harvard University (1974).

\bibitem{Phythian75}
R.~Phythian.
\newblock {\em J. Phys. A} {\bf 8}, 1423 (1975).

\bibitem{Phythian76}
R.~Phythian.
\newblock {\em J. Phys. A} {\bf 9}, 269 (1976).

\bibitem{Phythian77}
R.~Phythian.
\newblock {\em J. Phys. A} {\bf 10}, 777 (1977).

\bibitem{Andersen00}
H.~C. Andersen.
\newblock {\em J. Math. Phys.} {\bf 41}, 1979 (2000).

\bibitem{Eyink96a}
G.~L. Eyink.
\newblock {\em Phys. Rev. E} {\bf 54}, 3419 (1996).

\bibitem{Eyink96b}
G.~L. Eyink.
\newblock {\em J. Stat. Phys.} {\bf 83}, 955 (1996).

\bibitem{BeHo05}
A.~Berera and D.~Hochberg.
\newblock {\em Phys. Rev. E} {\bf 72}, 057301 (2005).

\bibitem{BeHo07}
A.~Berera and D.~Hochberg.
\newblock {\em Phys. Rev. Lett.} {\bf 99}, 254501 (2007).

\bibitem{BeHo09}
A.~Berera and D.~Hochberg.
\newblock {\em Nuclear Phys., B} {\bf 814}, 522 (2009).

\bibitem{Deker79}
U.~Deker
\newblock {\em Phys. Rev. A} {\bf 19}, 846 (1979)

\bibitem{Schwinger48}
J.~Schwinger.
\newblock {\em Phys. Rev.} {\bf 74}, 1439 (1948).

\bibitem{Dyson49a}
F.~J. Dyson.
\newblock {\em Phys. Rev.} {\bf 75}, 1736 (1949).

\bibitem{Zakharov75}
V.~E.~Zakharov and V.~S.~L'vov.
\newblock {\em Radiophys. Quantum Elec.} {\bf 18}, 1084 (1975).

\end{thebibliography}
\end{document}